\documentclass[11pt]{article}
\usepackage[utf8]{inputenc}
\usepackage{longtable}
\usepackage{subcaption}
\usepackage{graphicx}
\usepackage{placeins}
\usepackage{float}
\usepackage{amsmath, bbm, amssymb}
\usepackage[dvipsnames,table,xcdraw,svgnames]{xcolor}
\usepackage[hmargin=0.8in,vmargin={1.1in,1.1in}]{geometry}
\usepackage{hyperref}
\usepackage{multicol}
\hypersetup{
   colorlinks=true,
    linkcolor=blue,
   citecolor=blue,      
   urlcolor=blue,
}
\usepackage{natbib}
\setcitestyle{citesep={;}}
\usepackage[font={footnotesize}]{caption}
\usepackage{url}

\usepackage{parskip}
\usepackage{array}
\newcolumntype{P}[1]{>{\centering\arraybackslash}p{#1}}

\title{Mental health concerns prelude the Great Resignation: Evidence from Social Media\\
}

\author{R. Maria del Rio-Chanona$^{1,2\dagger}$, Alejandro Hermida-Carrillo$^{3,\dagger}$,  \\Melody Sepahpour-Fard$^{4,5}$, Luning Sun$^{6}$,   Renata Topinkova$^{7,8}$, and Ljubica Nedelkoska$^{1,2}$\\
\\
\footnotesize{$^{1}$ Complexity Science Hub, Vienna}\\
\footnotesize{$^{2}$ Harvard Growth Lab}\\
\footnotesize{$^{3}$ LMU Munich School of Management}\\
\footnotesize{$^{4}$ Science Foundation Ireland Centre for Research Training in Foundations of Data Science}\\
\footnotesize{$^{5}$ Department of Mathematics and Statistics (MACSI), University of Limerick}\\
\footnotesize{$^{6}$ The Psychometrics Centre, University of Cambridge}\\
\footnotesize{$^{7}$ Institute of Sociology of the Czech Academy of Sciences}\\
\footnotesize{$^{8}$ Faculty of Arts, Charles University}\\
\footnotesize{$^\dagger$ These authors contributed equally}\\
}
\date{\today}

\begin{document}

\maketitle

\begin{abstract}
To study the causes of the 2021 Great Resignation, we use text analysis to investigate the changes in work- and quit-related posts between 2018 and 2021 on Reddit. We find that the Reddit discourse evolution resembles the dynamics of the U.S. quit and layoff rates. Furthermore, when the COVID-19 pandemic started, conversations related to working from home, switching jobs, work-related distress, and mental health increased. We distinguish between general work-related and specific quit-related discourse changes using a difference-in-differences method. Our main finding is that mental health and work-related distress topics disproportionally increased among quit-related posts since the onset of the pandemic, likely contributing to the Great Resignation. Along with better labor market conditions, some relief came beginning-to-mid-2021 when these concerns decreased. Our study validates the use of forums such as Reddit for studying emerging economic phenomena in real time, complementing traditional labor market surveys and administrative data.
\end{abstract}


\section{Introduction}
In April 2021, 3.9 million workers quit their jobs in the U.S., the highest recorded quit rate in at least the last 30 years \citep{davis2014labor,BLS2021JOLTS_1}. During 2021, the U.S. quit rates remained high \citep{BLS2021JOLTS_1} and high quit rates were also reported by the media in other countries \citep{Horowitz_2022,Gupta_2021,Khan_2021}. This phenomenon has been named the ``Great Resignation" and received considerable attention from news outlets \citep{Casselman_2022,Rosalsky_2021,Donegan_2021,Romm_2021}. Given that people deeply care about employment events \citep{alan2011imperfect}, the record high quit rate raises concerns about worker well-being, and given the high costs of losing workers \citep{o2007cost}, the Great Resignation poses a big challenge for recovering businesses. 

In spite of the relevance of this topic, there is a dearth of academic literature studying this phenomenon. While some academics have discussed the topic narratively \citep{tessema2022great}, to our knowledge there are no academic articles exploring the causes of the Great Resignation in a data-driven manner. Nonetheless, there are several newspaper and magazine articles empirically covering the topic. A Wall Street Journal article \citep{wsj2022}, based on a survey, suggested that the pandemic caused some workers to drop out of the labor force permanently. Economist Paul Krugman analyzed U.S. labor market data and argued in the New York Times \citep{nyt2022}, that the rise in quits was less driven by labor force drop-outs and more due to people switching jobs and starting new businesses. 
\cite{cook2021who} and \cite{sull2022toxic} analyzed private data sets of employee profiles and records and identified mental health, burn-out, postponed resignations, and toxic work environments as possible explanations. Although these  newspaper and magazine articles present compelling causes of the Great Resignation, there is no consensus on the extent to which it was driven by job switching, self-employment, labor force reductions, burn-out and/or mental health concerns.

While the academic literature on the Great Resignation is scarce, there is an extensive literature in management science and labor economics studying quit behaviour from which we can draw upon. We know that people usually quit their jobs to pursue new, likely better ones \citep{halllazear1984,lazear2012hiring,lee2008understanding}. This leads to a pro-cyclicality in quit rates, they peak in economic expansions, when job openings are plentiful and of higher quality, and plummet in economic recessions, when job openings are scarce \citep{davis2014labor}. Most likely, the pro-cyclicality of quit rates is partially the story behind the 2021 Great Resignation. After a historically sharp disruption of the labor market in the first two quarters of 2020, the economy quickly recovered, creating record numbers of job openings, increasing nominal wages \citep{furman2022}, inducing job switches \citep{parkerhorowitz2022} and hence a high number of quits. In this sense, the pro-cyclical behaviour of job openings are \textit{pull factors} that increase the quit rate after a recession.  

However, it is unlikely that the pro-cyclicality of the quit rates explains the full extent of the Great Resignation. First, the increase in the labor market tightness (i.e., job openings to unemployed workers ratio) suggests a significant decline in matching efficiency between job-seekers and job openings \citep{rogers2022beveridge}. Second, the COVID-19 pandemic unleashed a chain of \textit{push factors} (i.e. factors that affected people's work experience and may have incentivized them to quit) not present in a typical recession that presumably contributed to the rising number of quits. In addition to the immediate consequences of personal exposure to the SARS‑CoV‑2 virus, workers had to cope with school closures and online schooling, caring after sick family and friends, and workplace burnout in sectors seen as essential. As a combined effect of the pandemic’s aftermath, working people experienced psychological pressures both at the workplace \citep{sull2022toxic, cook2021who}, and at home \citep{Donegan_2021}, and may have driven part of the recorded increases in anxiety \citep{ashokkumar2021social} and a worsening of the population’s mental health state \citep{xiong2020impact}. From a management science perspective, the COVID-19 pandemic was a shock (i.e., a jarring event
), and as such, may have triggered people to think about quitting \citep{lee1994alternative, morgeson2015event,akkermans2020covid}. These cognition processes have been labelled ``pandemic epiphanies" by the media \citep{Rosalsky_2021} and are also push factors the pandemic released.

Although the above mentioned push factors likely contributed to the rise in quits during the Great Resignation, it is difficult to identify these contributions through traditional economic data for several reasons. First, traditional labor market surveys and administrative data seldom include measures of mental health or other factors that may influence people's willingness to work. Second, some of the above cited articles  \citep{parkerhorowitz2022,sull2022toxic} rely on snapshot data from a single survey, making it difficult to capture pre- and post- pandemic changes. Third, surveys may limit participants' expression by having a set of predefined questions and, since employment can be a sensitive topic, the responses may be biased \citep{tourangeau2000psychsurveys}. Fourth, surveys and administrative data are costly and time-consuming to collect. The latter may be one of the main reasons why the Great Resignation is still under-explored in the academic literature.

In this paper, we analyze the content of work- and quit-related posts on Reddit from a majority of U.S. based users between 2018 and 2021 to measure changes in the work discourse and how it might relate to the rise in quits during the Great Resignation. We use data from Reddit, an online platform that allows users to discuss and share experiences around topics of interest called subreddits. These data poses an alternative for studying the Great Resignation overcoming the above-mentioned challenges. Reddit has the advantage that posts are semi-anonymous and of unrestricted length, allowing individuals to express themselves freely and in detail. Furthermore, data can be collected in real-time through an API \citep{baumgartner2020pushshift}. Recent works have also used Reddit to study socio-economic phenomena. \cite{semenova2021reddit} and \cite{lucchini2021reddit} 
studied the 2021 increase in GameStop's shares through the subreddit ‘r/WallStreetBets’. \cite{biester2021understanding} analyzed how the COVID-19 pandemic affected individuals' activity in mental health forums. \cite{ashokkumar2021social} showed psychological shifts during the pandemic using Reddit and validated this findings with surveys. \cite{waller2021quantifying} studied polarization in Reddit and showed that the demographics of various subreddits coincide with those of their real-world counterparts. 

We ask the following research questions. (How) did the mood around quitting change in the aftermath of the pandemic? On the one hand, one could expect a better sentiment since unemployment benefits and COVID-19 stimulus payments may have created a cushion for those considering to quit. The increasing number of job vacancies may also have reduced concerns about finding a new job. On the other hand, pandemic challenges such as adapting to remote work, school closure or health issues, may have worsened the sentiment around quitting a job.

(How) did the work discourse chance since the onset of the pandemic and during the Great Resignation? One could expect that people talked more about remote work, job switching and health and mental health concerns. Can we identify such changes on Reddit? If so, can we determine if, relative to the general work discourse, some of these topics increased more among users thinking about quitting? Given that that quit-cognitions are a strong predictor of quitting \citep{rubenstein2018surveying}, by answering this last question we can identify possible drivers of the 2021 record high in quits.

To answer these research questions, we use sentiment analysis and topic modelling, two widely used automated text analysis methods. Sentiment analysis captures emotions, sentiment, and attitudes in texts, usually through a pre-defined lexicon. It has been used to study both financial markets and voters' opinions \citep{10.1145/2436256.2436274} and, more recently, to understand how the pandemic affected individuals' emotions and risk perception \citep{ashokkumar2021social,dyer2020public}. Topic modelling aims to discover latent thematics in texts without imposing pre-defined categories. It has been used to study several topics including corporate funding and social responsibility \citep{jaworska2018doing,tonidandel2021using}, media framing dynamics \citep{heidenreich2019media}, citizen demands  \citep{vidgen2020and}, parenting concerns  \citep{sepahpour2022mothers}, extremists ideologies \citep{perry2021cognitive}, and climate-change polarization \citep{farrell2016corporate}.  In this study we use the NRC emotion lexicon \citep{mohammad2013crowdsourcing} for sentiment analysis and a Structural Topic Model \citep{roberts2014structural} to study changes over time in the work discourse. To measures changes on the quit-related discourse relative to the more general work discourse, we use difference-in-differences methods.

We find that the sentiment analysis reveals a strong increase in negative sentiment among quit-related posts during the first quarter of the pandemic, but then returns back to pre-pandemic levels. In other words we find that during the first quarter of the pandemic there was a wave of quits of despair, but this wave was short lived and faded before the Great Resignation. When using topic modelling, we find that the Reddit discourse captures known work-related changes that the pandemic brought, namely the rise of remote work and decline in commuting. Furthermore, switching jobs, work-related distress and mental health topics increased their prevalence after the start of the pandemic. These findings are in line with studies showing that the COVID-19 pandemic worsened the mental health of the general population across the world (see \citeauthor{xiong2020impact} for a review) and with the discussions relating the Great Resignation with job switching and mental health concerns \citep{Krugman_2021,cook2021who,sull2022toxic}.

Our main finding is that the pandemic exacerbated the already growing mental health concerns among workers, and we show that such concerns became disproportionately present in the discourse of quit-related posts since the onset of the pandemic. Furthermore, posts about mental health and work-related distress are more likely to be about quitting. We argue that, while the increase in job vacancies and job switching were factors present in previous recovery periods, the COVID-19 pandemic unleashed forces leading to quit behavior, such as mental health concerns, that were absent in previous recoveries. These additional factors could help explain the unusually high rates of quitting in 2021. 

To complement the above result we ask to what extend was the contribution of mental health concerns to increase in the quit discourse driven by (a) an increase in mental health concerns or (b) and increase in the strength of the relationship between mental health concerns and quits. That is, people may have been more likely to mention quitting when talking about mental health after the pandemic, regardless of whether more people talked about mental health or not. We use a multiple regression analysis to answer this question and find that the relationship between mental health concerns and quits remained roughly constant. This means that the onset of the pandemic did not change the character of the relationship between push factors and quit considerations, but it elevated the prevalence of phenomena such as mental health and work-related distress, which in return led to more quits. 

This paper is organized as follows. Section~\ref{sec:reddit_great_resignation} gives an overview of the Reddit population we study and how the share of quit- and fired- related posts resemble the quit and layoff rates of the U.S. In section~\ref{sec:sentiment} we present the sentiment analysis results. The results on topic modelling are divided in three subsections. First, section~\ref{sec:topic_modelling_1} we briefly explain the Structural Topic Model we use, describe the topics we focus the discussion on, and discuss how the general work discourse changed on Reddit after the start of the pandemic. Second, in section~\ref{sec:topic_modelling_2} we present our main finding, where we use a difference-in-differences analysis to understand how the quit-related discourse changed relative to the general work-related discourse. Third, in section~\ref{sec:topic_modelling_relationship} we present the multiple regression analysis discussed in the last paragraph.  
We discuss our results and their implications in section~\ref{sec:discussion} and for an overview of the methodology we refer the reader to the Methods section at the end.

\section{Results}

\subsection{Reddit and The Great Resignation}\label{sec:reddit_great_resignation}
We use data from the subreddit `r/jobs', a popular work-related subreddit targeted to a general working-age audience. As we discuss in the Supplementary Material~\ref{sec:SI_rjobs}, `r/jobs' has a posting history going a few years back before the pandemic, and is therefore better suited to study the causes of the Great Resignation than subreddits such as `r/antiwork', that have been recently covered by the news \citep{Rogers_2022}. 

Our unit of analysis are monthly time-stamped `r/jobs' posts from 2018-2021. We do not have panel data, that is, we do not follow Reddit users over time. Instead our underlying assumption is that the posts within each time interval resemble the work discourse of the general population. We use keywords to label posts as quit- and nonquit- related. We find 26,016 and 172,065 quit- and nonquit- related posts respectively.  To study how the quit-related discourse changed relative to the general work discourse, we build a control group of equal size of posts using a random sample of the nonquit-related posts. For the sentiment analysis we also filter posts of less than 15 words. This gives a corpus of slightly more than 50,000 posts, which we study through sentiment analysis and topic modelling. For a detailed description on how we label posts and build the control group we refer the reader to the \ref{sec:methods} section.

To contextualize our findings, we first study the demographic characteristics of the population of `r/jobs'. Measuring these characteristics is challenging; Reddit users are not geolocated, often use gender-neutral pseudonyms, and rarely portray personal information in their profile. Nonetheless, we can obtain some information about the `r/jobs' population by leveraging on a recent study by \cite{waller2021quantifying}, which used neural-embeddings to characterize the populations of various subreddits in terms of age, gender and U.S. political partisanship. Additionally, we provide some estimates by manually examining the posting history of a subsample of 400 `r/jobs' users to infer their demographic characteristics using an approach similar to \cite{von2021behavioral}. 

Our results (shown in detail in the Supplementary Material~\ref{sec:SI_reddit_population}) suggest that the sample of Reddit users we study is composed 
mostly of working age young adults that are based in the U.S. After the onset of the pandemic, the female population increased from 51\% to 68\%. The user population also became more international, though the U.S. remained by far the most common country (the share of U.S. users changed from 81\% to 70\%). When interpreting these estimates, we must take into account that they are based on self-disclosed information and may contain biases. Therefore, we use these estimates to provide context, but do not incorporate them into subsequent statistical analysis.

Considering that our sample of study is mostly U.S.-based and that a considerable amount of the Great Resignation media attention has been around this country, we focus our discussion within the context of the U.S. Note that we have the caveat that we cannot exclude users outside of the U.S. from our sample since it is infeasible to manually extract the geolocation of all users. Nonetheless, as we discuss in the following paragraphs, the evolution of the `r/jobs' discourse resembles the dynamics of the U.S. labor market. 

\begin{figure}[h]
    \centering
\includegraphics[width = .49\textwidth]{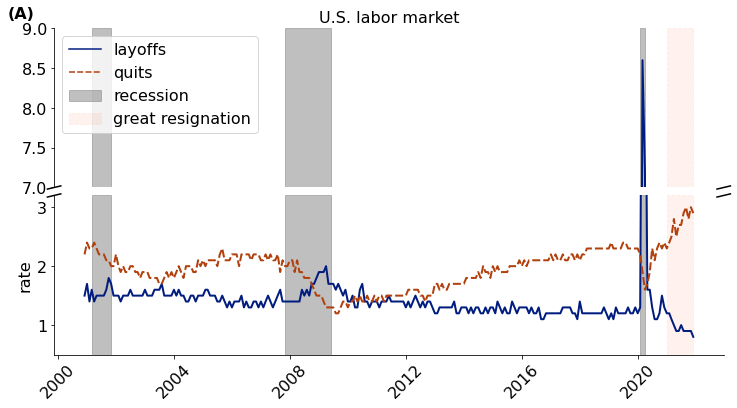}
\includegraphics[width = .49\textwidth]{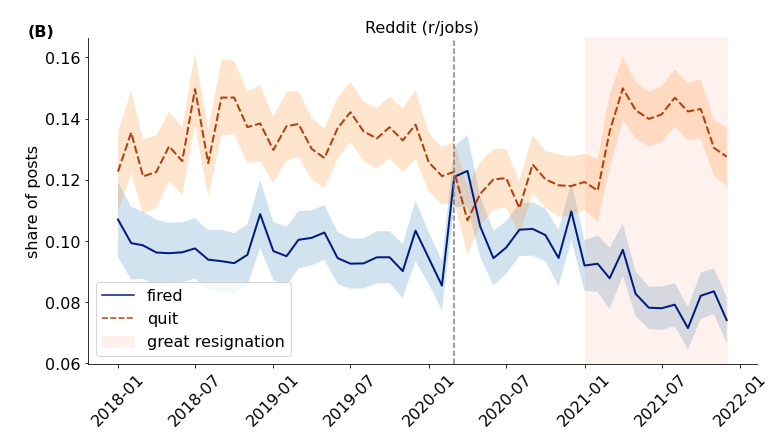}
    \caption{\textbf{US labor market and `r/jobs'} (A) The U.S. quit and layoff rates from December 2000 to December 2021. Recession periods are marked with grey shaded areas. (B) The share of quit- and fire- related posts from 2018 to 2021. In both panels the dashed vertical line corresponds to March 2020, the orange shaded area to the Great Resignation period, and the frequency is at a monthly level.}
        \label{fig:labor_market_summary}
\end{figure}

Figure~\ref{fig:labor_market_summary}A shows the U.S. quit and layoff rates from 2001 to 2021, portraying the record high quit rates in 2021 and its pro-cyclical behaviour. There is no clear start of Great Resignation, but we know it is a 2021 phenomena, hence we highlight with an orange shaded area the year 2021 as the approximate period of the Great Resignation in the plots of this paper. As discussed in the introduction, the pro-cyclical behaviour of the quit rate contributed to the Great Resignation, but is unlikely to be the full story. In the Supplementary Material~\ref{sec:SI_economics} we argue this further by showing that the relationship between job openings and quits has weakened in the U.S. In said section, we also include a more detailed description and discussion of the U.S. labor market and the Great Resignation using administrative data. 

To compare the dynamics of `r/jobs' discourse and the U.S. labor market, we compare the evolution of the share of quit-related posts with the quit rate. We also label posts as fired-related using keywords (see methods section for details) and compare the share of fired-related posts with the layoff rate. As shown in Figure~\ref{fig:labor_market_summary}, two years before the pandemic the shares of quit- and fired-related posts were roughly constant, similar to the rates in the U.S. labor market. In March and April 2020 the share of fired-related posts spiked and so did the layoff rate. In contrast, during this period, the quit share of post and quit rate decreased. In 2021 the share of fired-related posts decreased constantly, while the share of quit-related posts increased rapidly and then remained roughly constant. These dynamics also match the fact that the quit rate increased in 2021 and the layoff rate decreased. For a quantitative comparison between the time series we calculate the correlation between the time series. The correlation between the Reddit and the U.S. economy quit time series is $0.57$ (p-value $2.5\times 10^{-5}$), and the correlation between the layoff/fired time series is $0.60$ (p-value $5.5\times 10^{-6}$). In order words, quantitatively the dynamics of `r/jobs' are similar to the dynamics of the U.S. labor market. 

In the Supplementary material~\ref{sec:SI_rjobs} we present an additional analysis using posts' flairs (i.e., predefined tags users can add to specify a post's topic) that shows that the `r/jobs' discourse also qualitatively matches other aspects of the U.S. labor market such as an increase in job offers in 2021.

In this section we showed that the dynamics of the `r/jobs' resemble those of the U.S. labor market before and during the pandemic and at the start of the Great Resignation. This finding provides validation to our study and suggests that by investigating the change in the text content of the `r/jobs' posts we can better understand the Great Resignation. Later, in section \ref{sec:topic_modelling_1}, we provide additional validation by showing that with the start of the pandemic, topics related to working from home emerged and the discourse around commuting declined. For now, we turn to answer our research question about whether sentiment changed after the pandemic started and during the Great Resignation.

\subsection{Sentiment before and during the pandemic and the Great Resignation}\label{sec:sentiment}

Did the pandemic and the Great Resignation change the sentiment around job quitting, and if so, in which direction? To answer this question we look at changes in the sentiment score of the `r/jobs' posts. 
We use the NRC emotion lexicon \cite{mohammad2013crowdsourcing}, which identifies sentiment across polarity (positive and negative) and the eight basic emotions (fear, anger, sadness, disgust, joy, trust, surprise, and anticipation) according to the theory by \citet{plutchik2001nature}. We use the NRC lexicon \cite{mohammad2013crowdsourcing} since it has shown better performance in word-emotion lexicons than other approaches \citep{kuvsen2017identifying} (In the Supplementary Material \ref{sec:SI_sentiment} we check the robustness of our results with respect to other methods). 

First, we explore the overall time trends of the NRC sentiments for the quit- and nonquit- related posts together (see Figure~\ref{fig:sentiment_all_quarters} in the Supplementary Material). We find that in the two years leading to the pandemic the mood recorded on `r/jobs' was deteriorating. Negative emotions such as fear, sadness and disgust were trending upwards, while joy and trust were downward trending. The early months of the pandemic show record high levels of negative, and record low levels of positive polarity. However, we also see a more recent reversal in these trends. Fear and anger have been declining since the beginning of 2021, coinciding with the Great Resignation period, while joy and trust have stabilized since mid 2020. 

The more interesting question for us is, how did the sentiment among quit-related posts change compared to the control group of nonquit-related posts? To measure this, we use a difference-in-differences approach, where the treatment group is the quit-related posts. We use the control group to account for changes in sentiment that the pandemic may have caused across general posts in `r/jobs'. We use the following regression model
\begin{equation}
    y_{it}=\alpha + \beta_1Q_i+\sum \limits_{q=-7}^{7}\beta_2^qT_t^q + \sum \limits_{q=-7}^{7}\beta_3^qQ_iT_t^q +\epsilon_{it},
    \label{eq:DiD_sentiment}
\end{equation}
where $y_{it}$ is the sentiment of post $i$ published at time $t$, and $t={2018q1, ..., 2021q4}$ is measured quarterly. $Q_i$ is a dummy variable that takes a value of 1 if the post is quit-related (treatment group) and 0 if it isn't (control group). $T_t^q$ is a set of dummies corresponding to 15 quarters, seven pre-pandemic, one tagging the start of the pandemic (Q1, 2020), and 7 during the pandemic. Each takes a value of 1 for the specific quarter that they indicate, and zero otherwise. The very first quarter (Q1, 2018) is left out of the equation. 
The coefficient $\beta_1$ shows the overall difference in sentiment between the treatment and the control group, while the vector of coefficients $\beta_3^q$ shows the differences between the groups over quarters ($q$). The equation also includes a constant $\alpha$, an error term $\epsilon_{it}$, and a set of time dummies $T_t^q$ that capture the overall time trend in a given sentiment and their coefficients $\beta_2^q$.

\begin{figure}[h]
    \centering
\includegraphics[width = 0.95\textwidth]{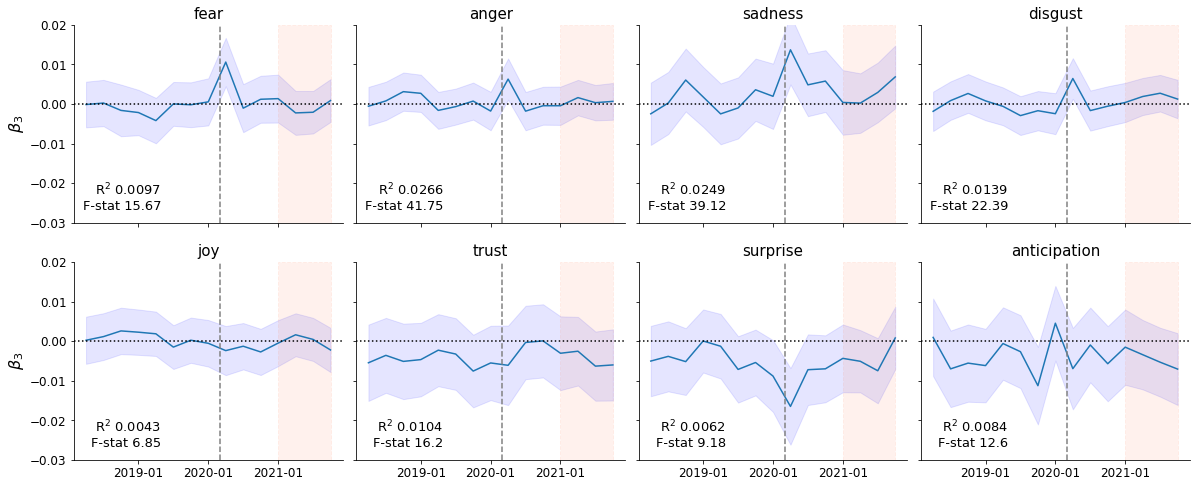}
    \caption{\textbf{Sentiment analysis} This plot shows the interaction coefficients $\beta_3$ for each quarter. The vertical line corresponds to the start of the pandemic (March 2020) and the orange shaded area indicates the approximate period of the Great Resignation (year 2021). Below each sub-graph we show the adjusted R-squared and the F-stat of the regression model.}
    \label{fig:sentiment_quit_quarters}
\end{figure}

We focus our analysis on the basic emotions, since the difference-in-differences analysis for the NRC polarity scores have high volatility and do not present any significant trends (see Figure~\ref{fig:app_sent_posneg} in the Supplementary Material). Figure~\ref{fig:sentiment_quit_quarters} shows that negative emotions (fear, anger, sadness and disgust) all spiked among the quit-related posts relative to the control group in the $2^{nd}$ quarter of 2020, but this effect was short-lived. We do not find any notable relative changes in emotions among quit-related posts since the start of the Great Resignation. In the Supplementary Material~\ref{sec:SI_sentiment} we use two other sentiment analysis methods and also find an increase in negative sentiment during the first quarter of the pandemic. This increase is short lived, and negative sentiment returns to pre-pandemic levels within two quarters after the start of the pandemic. 

The quarter in which negative sentiment among job-quitters spiked is largely an outlier in terms of low quit rate, low rate of job openings, and high rate of layoffs in the U.S. economy (see Supplementary Materials~\ref{sec:SI_economics}). Those quitting at this time, in the midst of the worst labor market they have experienced in their lifetime, seem to be quitting out of despair. 
Hence, our results suggest that at the start of the pandemic there was a wave of quits out of despair for a quarter of a year. Notwithstanding, as shown by the U.S. quit rate, this wave had a much smaller number of quits relative to the numbers observed during the Great Resignation that followed.

Although the results obtained by sentiment analysis are informative and reveal how people quitting at the start of the pandemic were in dispair, it is difficult to detect changes in narrative with this method. The sentiment analysis is a dictionary based approach agnostic to the topic being discussed. While it is useful to understand shifts in the emotional and sentiment load of texts, it would not detect changes in narratives when these are not accompanied by alterations in the emotions expressed. Furthermore, usually several topics change prevalence over time, and the changes in sentiment of each topic may average out within the overall sentiment score. To understand how the work discourse changed and identify possible drivers of the Great Resignation, we need a more refined method that infers the content of posts directly from the text without imposing pre-defined categories. To do this, in the next sections, we study the `r/jobs' discourse using a Structural Topic Model.

\subsection{Changes in the work discourse during the COVID-19 pandemic and the Great Resignation}\label{sec:topic_modelling_1}
Here we study how the topics in `r/jobs' posts changed since the start of the pandemic and relative to the pre-pandemic period. For example, were users more likely to talk about remote work, compensation or mental health after the onset of the pandemic? To quantify changes in the work discourse we use a Structural Topic Model~\citep{roberts2014structural} that allows us to document what individuals talk about when they make a work related and quitting-related post. In a Structural Topic Model, topics are a distribution over words and documents are a distribution over topics. The \textit{prevalence} of a topic in a document is the probability that a document belongs to a said topic.

\paragraph{The Structural Topic Model and description of important topics}  

The Structural Topic Model we fit to our corpus (made of both the quit and control groups of posts) has 90 topics; 68 of them are \textit{clear topics} in the sense that there is a clear identifiable theme, 10 of them are \textit{multi-topics} and include posts within two or three themes. The 12 remaining topics are noisy topics with difficult-to-identify themes; these topics are often referred to as \textit{boiler-plate} topics \citep{dimaggio2013exploiting}. In our analysis, we include only clear topics and multi-topics (78 topics in total). For brevity, we shorten multi-topics names to one or two theme and mark them with a star at the end of the name. Table~\ref{tab:si_topc} in Supplementary Material~\ref{sec:SI_topicmodelling} reports the full name and type of all topics, as well as their most frequent and exclusive (FREX) terms. When interpreting changes in prevalence of multi-topics we nuance the findings since the changes could arise from changes in any topic or combination of topics. In the methods section~\ref{sec:methods} we describe our methodology for fitting the Structural Topic Model and labelling topics; in the Supplementary Material~\ref{sec:SI_topicmodelling} we show robustness tests for our model fit. 

Several of the 78 interpretable topics are relevant within the context of the COVID-19 pandemic and the Great Resignation. In the following paragraphs we explain what these topics are about and discuss their dynamics through the following sections. The topics \textit{working from home} and \textit{remote jobs} include discussions about issues and anecdotes of remote working conditions, as well as searching for remote jobs. \textit{Remote jobs} is mostly about searching for remote jobs,  while \textit{working from home} is more about narrating working from home experiences, jobs or related issues. In contrast to these topics, the topic \textit{community, moving for a job} includes worries about taking a job that is far from home and discussions of whether it would be good to relocate to reduce commuting time.

Topics about health issues include \textit{Work-related distress},  \textit{mental health}, and the multi-topic \textit{health issues / healthcare job / scheduling} (from here onward referred to as \textit{health issues*} for brevity). \textit{Work-related distress} and \textit{mental health} are similar topics that are centered around negative psychological experiences and discuss anxiety, stress and depression -- the three mental illnesses most discussed in workplace settings \citep{harvey2017can,joyce2016workplace}. The former topic captures job-related distress and includes posts about feeling lost, stressed, anxious and/or overwhelmed at work. In the latter topic, posts discuss mental health concerns and explicit psychological disorders such as depression and anxiety.  \textit{health issues*} includes worries about health, experiences by healthcare workers, and general work scheduling issues. These topics seem to be clustered together since discussing health issues can include requesting/scheduling appointments (\textit{request} is included in the topic's top 10 terms), which is also a work management issue. This term also pops-up together when discussing healthcare jobs, as these are characterized by work shifts. As explained previously, we nuance findings when interpreting the results of this multi-topic due to identification issues.

We also find topics related to job quitting intentions and experiences, such as \textit{switching jobs, guilt about leaving for a better job, quit, resignation letters} and \textit{quitting a new job}. Two topics stand out in the context of the Great Resignation. In \textit{switching jobs} people discuss wanting to switch their job and questions about how to handle a job switch. The other topic that stands out is \textit{hate job \& want to quit}, which includes narratives of employees fed up with their jobs, among others, due to toxic workplaces (one of the FREX terms). Toxic workplaces and job switching are two of the proposed drivers of the Great Resignation \citep{sull2022toxic,nyt2022}. Not related to quitting, but related to toxic work, the topic \textit{hating job}, where people mention they hate or strongly dislike either their co-workers or the the tasks they do in their job.

There are topics related to earnings and careers. For example, \textit{salary negotiations}, where people ask how to negotiate salary in a job offer or ask for a raise in their current job, \textit{online jobs to make extra money}, which is about side jobs, mostly remote, to earn extra money, and \textit{make money}, where users, many of them teenagers, ask about how they can obtain some earnings. Two other topics related to people's careers are \textit{job titles, promotions}, where users ask about changes in their job title and/or mention they got a job promotion; and \textit{job offers issues}, which include posts about not hearing back after accepting a job offer, but also asking for advice about what to do when the user accepted one job offer, but then got a better one. 

With respect to job searching, the most general related topic is \textit{job searching}, where people report their experiences on searching for a job and ask advice. There are also two other self-explanatory topics: \textit{online job search} and \textit{difficulty finding a job}. One last topic related to job searching is the multi-topic \textit{looking for jobs / sales related questions} (from here onward referred to as \textit{looking for jobs*} for brevity), which is includes posts about looking for jobs mostly on sales and questions related to sales.

\paragraph{Changed in the work discourse}
To understand how the work discourse changed with the pandemic, we estimate a regression model for the prevalence of each topic ($y_i$) across documents and time (i.e. for the document-topic prevalence matrix). As a first approach, we measure the average difference in prevalence of topics before and after the onset of the pandemic. Later on we explore temporal variations in more detail. For the first approach we use the following regression
\begin{equation}
    y_{i} = \alpha + \beta_1T_{t} + \beta_2Q_{i}+\epsilon_{i}
    \label{eq:DiD_topic_basic},
\end{equation}
where $\beta_1$ is the vector with the coefficients of interest, which measure changes in prevalence in time for each topic.  $T_t$ are dummy variables which take a value of 0 until February 2020, and a value of 1 starting in March 2020; $Q_i$ are control variables, controlling for the structure of the sample (quit- vs. nonquit- related posts), and take a value of 1 for posts that are quit-related, and 0 for posts in the control group; $\alpha$ are constants and  $\epsilon_i$ the error terms. In the next section we use a regression analysis

\begin{figure}[h!]
    \centering
     \includegraphics[width = .5\textwidth]{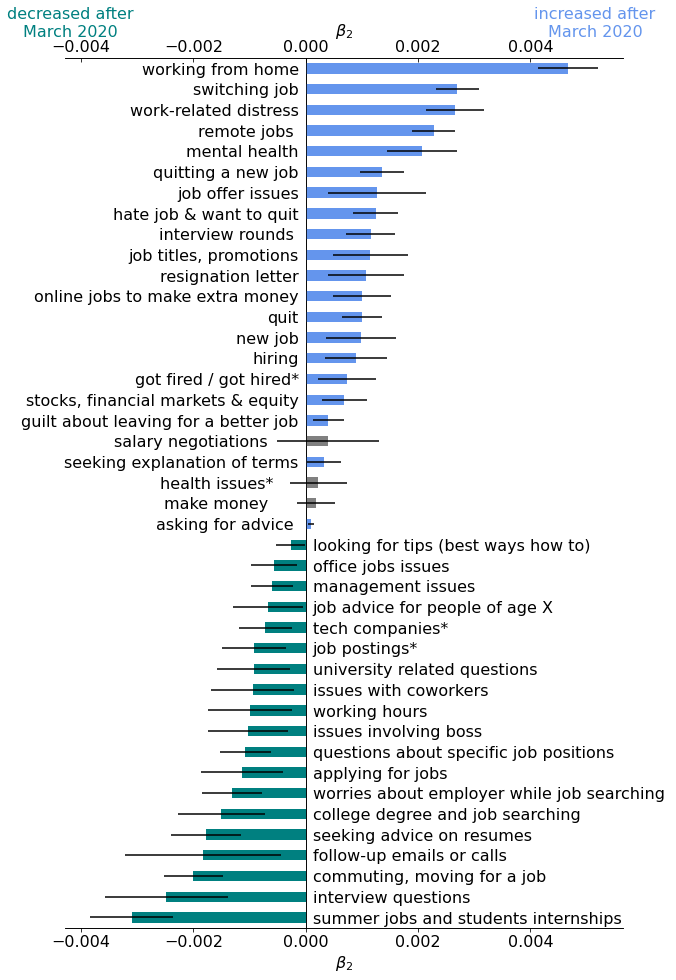}
    \caption{\textbf{Changes in overall topic prevalence} This figure shows the time coefficient of the regression in Eq.~\ref{eq:DiD_topic_basic}. Blue bars indicates a positive coefficient with a p-value below 0.05, while green bars indicate a negative coefficient with a p-value below 0.05. In grey we added three relevant topics that do not present a significant change (p-value above 0.05). Topics with a star at the end correspond to multi-topics and name may be shortened to a single topic. For full name of multi-topics please see Table \ref{tab:si_topc}.
    }
        \label{fig:STM_changes}
\end{figure}

 Figure \ref{fig:STM_changes} summarizes the changes in topic prevalence we estimated using Eq.~(\ref{eq:DiD_topic_basic}). This figure shows the coefficients $\beta_1$ from the regressions where the p-values are significant (below 0.05). 39 out of 78 topics changed their average prevalence significantly after the onset of the pandemic. Topics that increased their prevalence are colored in blue, while those that decreased in green. There are three additional topics colored in grey that did not present a significant change in the average prevalence but are of research interest in the context of the Great Resignation. Namely, we added the topics \textit{salary negotiations}, \textit{make money}, and \textit{health issues*}, as they are related to wages and health.

The topics with the largest increase in prevalence are: \textit{working from home}, \textit{switching jobs}, \textit{work-related distress}, \textit{remote jobs}, and \textit{mental health}. Moreover, five topics related to job quitting increased their prevalence significantly - \textit{quitting a new job}, \textit{hate job \& want to quit}, \textit{resignation letter}, \textit{quit}, and \textit{guilt about leaving for a better job}. Among the topics that show the greatest decline in prevalence we have \textit{commuting, moving for a job}. These changes resemble some known facts about how work changed due to the pandemic (e.g., the rise of remote work and decrease in commute \citep{brynjolfsson2020covid}) and the Great Resignation (e.g., an increase in quit rates).

We perform a structural break Wald test to understand whether the pandemic caused the change in topics' prevalence. This test determines whether the coefficients of the linear trend fitted to each topic time series differ before and after the pandemic. Therefore, this test allows us to discard changes due to a pre-existing upward or downward trend. 
We find that the five topics with the largest increase in prevalence show significant structural breaks in Q1 of 2020 (p-value of 0.05 or lower). Except for the \textit{summer jobs and student internships} topic, the top four topics that decreased their prevalence the most also show a structural break in the Q1 of 2020. Other topics that declined in prevalence with a level break in Q1 of 2020 (significant at 0.05 or lower) are \textit{applying for jobs} and \textit{management issues}. These results suggest that it was indeed the pandemic that caused the changes in the prevalence of the topics most related to the Great Resignation. 

\begin{figure}[h!]
    \centering
     \includegraphics[width = .75\textwidth]{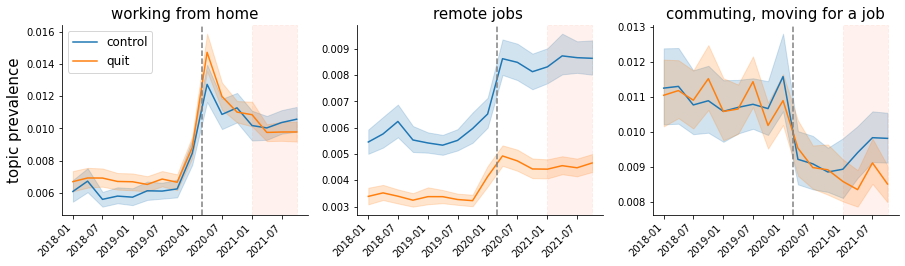}
    \caption{\textbf{Dynamics of topics related to working during the pandemic.} This plot shows the prevalence across time (in quarters) for the topics \textit{working from home}, \textit{remote jobs}, and \textit{commuting, moving for a job}. Quit- and nonquit- related posts are in orange and blue, respectively. The shaded areas around the time series denote the 95\% confidence intervals.  The dashed grey line marks the onset of the pandemic (March 2020) while the shaded area the period of the Great Resignation (2021). 
    }
        \label{fig:tm_timeseries_validation}
\end{figure}

\begin{figure}[h!]
    \centering
     \includegraphics[width = .75\textwidth]{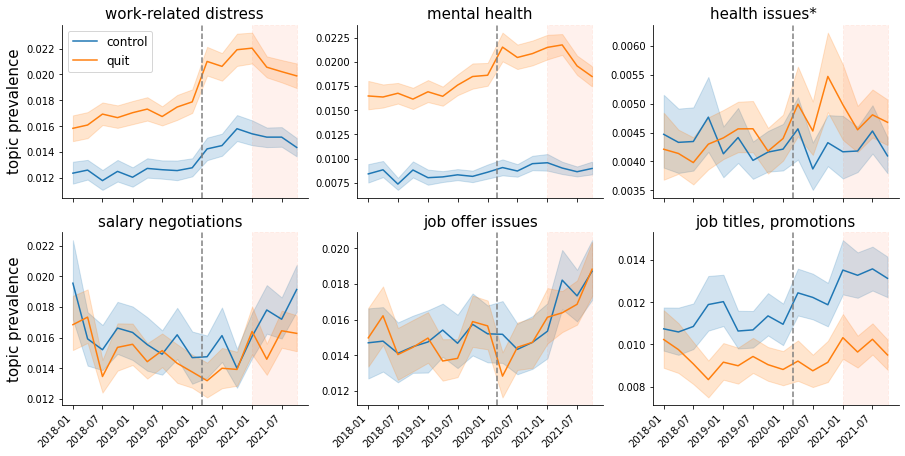}\\
     \caption{\textbf{Topics' prevalence changes across time} Prevalence across time (in quarters) for the topics \textit{work-related distress}, \textit{mental health}, and \textit{health issues/jobs in healthcare/scheduling} issues on the top row and \textit{salary negotiations}, \textit{job offer issues} and \textit{job title, promotions} on the bottom row. Quit- and nonquit- related posts are in orange and blue, respectively. The shaded areas around the time series denote the 95\% confidence intervals. The dashed grey line marks the onset of the pandemic (March 2020) while the shaded area the period of the Great Resignation (2021).
    }
        \label{fig:tm_timeseries_badeffects_positive}
\end{figure}

The labor market changed substantially between the first quarters of 2020 and 2021 - from record high fires to a record high quit rate - hence, it is also likely that the work discourse changed within this period. To analyze these finer-grained changes, we plot the quarterly dynamics of topics related to the pandemic and the Great Resignation. We split the topics across figures by putting together topics relating to working conditions that changed during the pandemic (Figure~\ref{fig:tm_timeseries_validation}), topics that reflect possible positive or negative effects of the pandemic and the Great Resignation (Figure~\ref{fig:tm_timeseries_badeffects_positive}), and topics about quitting (Figure~\ref{fig:tm_timeseries_quit}). For the dynamics of the rest of the topics please see Supplementary Material~\ref{sec:SI_topicmodelling}.

Figure \ref{fig:tm_timeseries_validation} shows that after the onset of the pandemic, the topics \textit{working from home} and \textit{remote jobs} increased their prevalence dramatically and that the prevalence of \textit{commuting, moving for a job} decreased sharply. While the prevalence of \textit{remote jobs} and \textit{commuting, moving for a job} remained roughly constant between 2020 and 2021, \textit{working from home} peaked immediately after the onset of the pandemic but decreased its prevalence shortly after. This latter pattern suggests that there was a period of sense-making, during which individuals were unsure about how to deal with the unexpected shift to the new work scheme. The decline in discussions around commuting and moving for a job mirrors the increase in discussions around remote work and working from home. The pandemic-related trends around remote work, home office, and commuting confirm what we already know from other studies \citep{brynjolfsson2020covid, mcfarland2020impact}, landing additional validation for using Reddit as a source of data to monitor the development of job-related attitudes.

(How) did the Great Resignation change the work discourse? Figure \ref{fig:tm_timeseries_badeffects_positive} shows that the prevalence of topics related to job opportunities increased during the Great Resignation; in contrast the prevalence of topics related to detrimental effects decreased. \textit{Salary negotiations}, \textit{job offer issues}, and \textit{job titles, promotions} increased their prevalence in 2021, resembling the improving labor market conditions for workers during the economic expansion. This point also comes across in Figure~\ref{fig:topic_prevalence_apx_1} in the Supplementary Material (fourth row), which shows that \textit{difficulty finding a job} decreased its prevalence in 2021. In the second half of 2021 the topics \textit{work-related distress}, \textit{mental health}, and \textit{health issues*} also decreased their prevalence, particularly among quit-related posts (we will examine this latter point more closely in the next section). The dynamics of the topics discussed here reflect how the labor market conditions improved in 2021 and suggest that these improvements may have given some relief to the detrimental effects of the pandemic.

The dynamics of posts related to quitting present a similar narrative. \textit{Switching jobs}, one of the topics that increased the prevalence the most, presents a sharp increase at the onset of the pandemic; afterward, the prevalence decreased but remained above pre-pandemic levels (see Figure \ref{fig:tm_timeseries_quit}). This finding supports what some economists have argued -- the Great Resignation is more of a Great Reshuffling with workers switching jobs rather than leaving the labor force \citep{nyt2022}. The prevalence of topics expressing negative feelings about quitting, such as \textit{hate job \& want to quit}, and \textit{quit}, increased sharply in the Q1 of 2020 -- this was a significant structural break with a Wald test with respective p-value below 0.05 -- and then decreased considerably by the start of 2021. In contrast, neutral topics topics about quitting, such as \textit{resignation letter}, and topics related to quitting for a new job, such as \textit{quitting a new job} and \textit{guilt about leaving for a better job}, increase their prevalence during 2021. Taken together, the changes in prevalence of topics presented in Figures \ref{fig:tm_timeseries_badeffects_positive} and \ref{fig:tm_timeseries_quit} suggest a change in mood around quitting during the Great Resignation -- from quitting out of despair in the pandemic to quitting for a better job opportunity.

\begin{figure}[h]
    \centering
     \includegraphics[width = .75\textwidth]{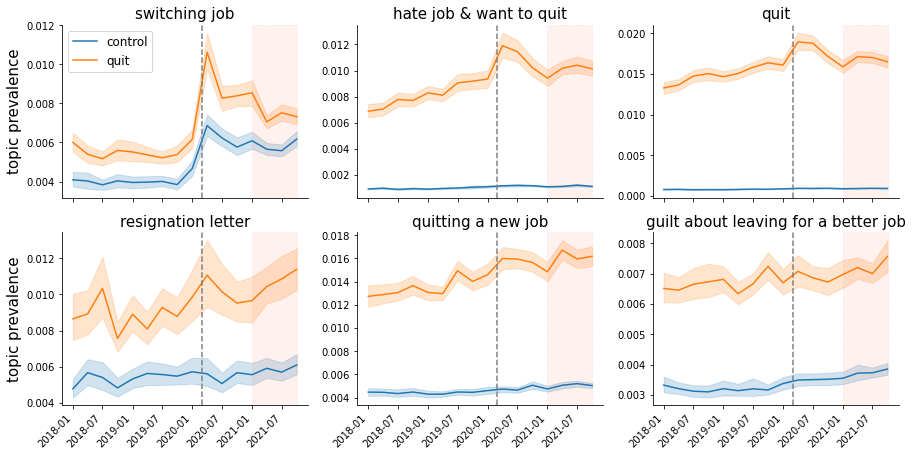}
  \caption{\textbf{Topics' prevalence changes across time} Prevalence across time (in quarters) for topics \textit{switching job}, \textit{hate job \& want to quit}, and \textit{quit} on the top row and \textit{resignation letter}, \textit{quitting a new job}, and \textit{guilt about leaving for a better job} on the bottom row. Quit- and nonquit- related posts are in orange and blue respectively. The shaded areas around the time series denote the 95\% confidence intervals.  The dashed grey marks the onset of the pandemic while the shaded area the period of the Great Resignation (2021) 
    }
        \label{fig:tm_timeseries_quit}
\end{figure}

The findings in this section complement those of the sentiment analysis. While quits were on the rise long before the pandemic, the pandemic intensified people's intentions to quit. The  topics around quitting changed both at the onset of the pandemic, and during the Great Resignation. The start of the pandemic induced  concerns about mental health, and work-related distress. However, some of these concerns were resolved when the Great Resignation started, likely due to better employment opportunities. People talked less about quitting because they hated their job, and more about negotiating salaries and discussing job offer issues. Some of the people quitting in 2021 might have strongly disliked their job in 2020, but did not quit until 2021, when there were better employment prospects. Others, likely the ones that felt guilty about quitting, might not have disliked their job but left it for a better employment opportunity. 

\subsection{Changes in the quit-related discourse and causes of the Great Resignation}\label{sec:topic_modelling_2}
Having documented how the general work related discourse changed in `r/jobs' after the onset of the pandemic, we turn our attention to specific changes in the quit-related discourse. To understand how the quit-related discourse changed not only relative to the pre-pandemic levels, but also relative to posts that are not quit-related (i.e., relative to the control group), we estimate an event study style difference-in-differences model as outlined in equation (\ref{eq:DiD_sentiment}). However, instead of sentiments, $y_{it}$ now stands for various topics identified by the Structural Topic Model. 

We present the results of the difference-in-differences analysis for the all 78 topics in Table \ref{tab:si_topc_did}. Here in the main text, we focus on the topics for which the difference-in-differences analysis satisfies the parallel trends assumption (until a one quarter pre-trend) and that are relevant in the context of the Great Resignation. 

\begin{figure}[h]
    \centering
    \includegraphics[width = 0.95\textwidth]{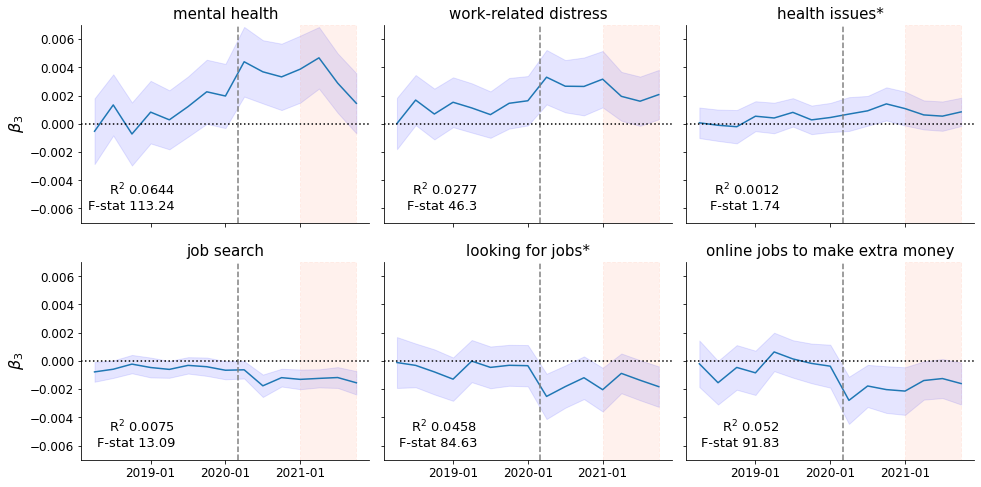}
    \caption{\textbf{Difference-in-Differences analysis for topics}. Relative changes in prevalence in selected topics among quit-related posts.  Positive values indicate an increase in prevalence and negative values a decrease. The dashed grey line marks the onset of the pandemic (March 2020) while the shaded area the period of the Great Resignation (2021). The frequency is quarterly.}
    \label{fig:did_topic_mainl}
\end{figure}

\textit{Mental health}, \textit{work-related distress} and \textit{health issues*} significantly increased their prevalence among quit-related posts relative to nonquit-related posts since the start of the pandemic. \textit{Work-related distress} increased in prevalence by 11.4 percent among nonquit-related posts since the pandemic, and by 17.5 percent among the quit-related ones. The average relative difference is 6 percentage points (pp) for the quit-related posts. \textit{Mental health} increased by 5.5 percent among the nonquit-related posts since the pandemic, and by 15.6 percent among the quit-related ones, a 10 pp difference. 

This finding portrays once again the strong distress that people who wanted to quit their job in 2020 were in. Towards the end of 2020 the relative difference in the prevalence of \textit{mental health} and \textit{work-related distress} decreased, likely reflecting the improved labor market conditions. Nonetheless, it remained higher than in the pre-pandemic period. We take this as suggestive evidence that, in contrast with the pre-pandemic years, since the pandemic, including the period covering the first year of the Great Resignation, people that are quitting are more concerned with mental health and emotional stress.

The topic \textit{health issues*} follows a similar pattern as \textit{mental health} and \textit{work-related distress}, but the relative change is smaller in magnitude. Its prevalence increased by 8.3 percent among the nonquit-related posts, and by 13.5 percent among the quit-related ones, an average difference-in-differences of over 5 pp. As discussed in the previous section \textit{health issues*} is a multi-topic. It includes posts related to patients issues such as scheduling and healthcare workers treating patients, among other things. Therefore we cannot clearly conclude what part of health issues drives the pattern, and we are unable to further disentangle the topics by, e.g., analysing the most common word-counts (see Supplementary Material~\ref{sec:SI_topicmodelling}). At most we show evidence that health issues were predominant in the quit-related discourse in the period of the pandemic. 

Another topic worth discussing in the context of work mental health and toxic work environments is \textit{hate job \& want to quit}. While this topic increased its prevalence among quit-related posts relative to the control group after March 2020, we cannot conclude the pandemic caused this increase since this topic has an upwards pre-trend. As shown in Figure \ref{fig:app_topic_toxic} and in Table \ref{tab:si_topc_did}, the prevalence of \textit{hate job \& want to quit} grew faster among quit-related posts pre-pandemic than in the general work discourse. This pre-pandemic increase was partly due to the word `toxic' being used more often (see Figure\ref{fig:app_topic_toxic}). Nonetheless, it still holds that in the first three quarters of the pandemic \textit{hate job \& want to quit} increased its relative difference in prevalence more sharply than before the pandemic. This difference then converged back to the pre-pandemic trend in 2021. These results suggest that, although the pandemic was not a root cause of increasing quits due to toxic work environments, it may have exacerbated this problem in 2020.

The difference-in-differences analysis reveals that people seem to be less worried about finding a new job when talking about quitting after the onset of the pandemic. As the bottom panels of  Figure~\ref{fig:did_topic_mainl} show, topics related to job searching, such as \textit{looking for jobs*}, \textit{online job search}, and \textit{job searching} decreased their prevalence in quit-related posts in comparison to the control group. Perhaps surprisingly, topics that portrayed a more positive status of the labor market (i.e., \textit{salary negotiations}, \textit{job titles, promotions}, and \textit{job offer issues}) did not show a significant difference between the treatment and the control group (see Table \ref{tab:si_topc_did}). This does not mean that these issues are not relevant for workers quitting their jobs, but that they changed equally for both groups.

Taken together our results show that, among people that were talking about quitting, mental health concerns increased after the onset of the pandemic and before the Great Resignation. 
This finding suggests that distressing experiences at work and concerns about work-related mental health may have increased people's motivation to quit. When more vacancies opened in 2021, some workers may have taken the opportunity to quit and switch jobs, as the increase in the prevalence of \textit{switching jobs} would suggest, or leave the labor force. The Great Resignation also seems to have brought some relief to mental health concerns, particularly towards the end of 2021, where we see mental health-related topics decrease their prevalence towards pre-pandemic levels.

\subsection{ The relationship between mental health concerns and quitting: a multiple regression model}\label{sec:topic_modelling_relationship}

Lastly, to better understand whether push factors such as mental health concerns, health issues, or toxic work environments are linked to quits, we measure the strength of the relationship between a post being quit-related and the prevalence of topics related to these push factors, while controlling for other topics discussed in `r/jobs' posts. To do this we use a sample of posts with the same composition of quit- and nonquit- related posts than the general `r/jobs' population of posts (see Methods sections for details). From the 78 interpetable topics we chose four topics that potentially identify push factors for quitting: \textit{mental health}, \textit{hating job}, \textit{work-related distress}, and \textit{health issues*}. We chose \textit{hating job} since it relates to toxic work culture and is similar to \textit{hate my job \& want to quit}, but is not confounded with posts being quit-related.
In addition, we identify factors that we think might also affect the probability of quitting. These are factors that usually improve the conditions of a job: (flexibility of) \textit{working from home}, \textit{salary negotiations} and \textit{job promotions}.  We choose \textit{working from home}, rather than \textit{remote jobs}, since as we explained previously, \textit{working from home} is more related to the experience and \textit{remote jobs} is more about users looking for remote jobs. 

We estimate the following model using a logistic regression
\begin{equation}
    Q_{it} = \boldsymbol{X_{it}^{'} \beta}+\boldsymbol{Y_{it}^{'} \gamma} + \boldsymbol{T^{'}_t \delta} + \varepsilon_{it},
    \label{eq: quits}
\end{equation}
where $Q_{it}$ is a binary variable indicating whether a post is quit-related $(1)$ or not $(0)$, $\boldsymbol{X_{it}}$ is the set of topics indicating push factors, whose set of coefficients $\boldsymbol{\beta}$ we are interested in. $\boldsymbol{Y_{it}}$ is the set of topics related to factors that may reduce one's interest in quitting the job. $\boldsymbol{T_t}$ is a full set of time dummies, indicating months, which control for economy-wide time-variant factors that also affect the rate of quitting, such as the rate of job openings, and other business cycle changes. We estimate the same model twice, once including all months available in the sample, and once only for the period since the start of the COVID-19 pandemic (March 2020 - December 2021). We do this in order to see if the pandemic changed the character of the relationship between the push factors and quit considerations, or whether it merely aggravated certain push factors.  
\begin{table}[h]
    \centering
    \caption{Explaining Quit Considerations}
\begin{tabular}{lcc} \hline
 & (1) & (2) \\
  & 2018-2021 & Pandemic \\ \hline
 &  &  \\
Mental health & 1.948*** & 1.919*** \\
 & (0.125) & (0.175) \\
Hating job & 1.291*** & 1.319*** \\
 & (0.0977) & (0.128) \\
Work-related distress & 0.385*** & 0.404*** \\
 & (0.0892) & (0.110) \\
Health issues & 0.0354 & 0.167 \\
 & (0.179) & (0.202) \\
Working from home & -0.456*** & -0.461*** \\
 & (0.143) & (0.168) \\
Salary negotiations & 0.163*** & 0.208*** \\
 & (0.0455) & (0.0644) \\
Promotions & -0.213* & -0.414** \\
 & (0.116) & (0.167) \\
 &  &  \\
 Log-likelihood constant only model & -10582.812 & -5275.789 \\ 
 Log-likelihood & -9976.873 & -4926.843 \\ 
 Pseudo R-squared & .057 & .066 \\ 
 Observations & 29,271 & 14,722 \\ \hline
    \multicolumn{3}{c}{ Robust standard errors in parentheses} \\
    \multicolumn{3}{c}{ Significant at: *** p$<$0.01, ** p$<$0.05, * p$<$0.1} \\
\end{tabular} 
    \label{tab:quits}
    
    \caption*{Note: The first model includes all months available in our sample (January 2018 - December 2021). The second model includes the months since the start of the COVID-19 pandemic (March 2020 - December 2021).}
\end{table}

Table \ref{tab:quits} shows the results of estimating equation (\ref{eq: quits}). For easier interpretation, the coefficients are expressed as marginal effects estimated at the mean of the other variables. The coefficients in the two models are similar, but the second model has a better fit (a higher log-likelihood and a higher pseudo R-squared). Hence we focus on the model fitted for the period since the start of the pandemic. 

The results suggest that a change in prevalence in the mental health topic from 0 to 1 in a post almost doubles the quit probability (i.e. the probability that the post is quit-related). However, topics rarely have prevalence close to 1 in a post, so another way to express the result would be to ask how the quit probability increases if we move from the $5^{th}$ to the $95^{th}$ percentile of post within which \textit{mental health} is a prevalent topic. Such move corresponds with an increase in the quit probability of 6.1 percent. Similarly, a change in prevalence from 0 to 1 in hating one's job is associated with 132 percent higher quit probability, and a move from the $5^{th}$ to the $95^{th}$ percentile of this topic corresponds with an increase in the quit probability of 5 percent. A similar move in the topic \textit{work-related distress} results in 1.7 percent higher quit probability.

The relationship between health issues and quits is not significant. This does not mean that health issues (that are not about mental health) did not play a role in explaining quits, but that 
we are not able to detect this effect. Most likely, this is because, \textit{health issues*} is a multi-topic that captures both personal health issues, issues raised by health sector personnel, as well as issues around scheduling general appointments. In other words, we do no have a homogeneous topic that can capture the effect of health issues.

Among the job-improving 
factors, \textit{working from home} appears as the strongest predictor of a post being quit-related. As expected, higher prevalence of this topic is associated with lower quit probability. Moving from the $5^{th}$ to the $95^{th}$ percentile of this topic corresponds with a decline in the quit probability of 1.5 percent. We find no significant effect of discussing \textit{job promotions} on the quit probability, while the relationship between \textit{salary negotiations} and quits is opposite of what one might expect. Moving from the $5^{th}$ to the $95^{th}$ percentile of this topic increases the quit probability by 1.4 percent. One reason why this relationship is positive may be that some of the salary negotiations discussed on `r/jobs' are about being denied a raise and hence the user may consider quitting. Another reason may be that users talk about a job offer they got with a better salary, and discuss whether they should quit their job or renegotiate their salary.

Importantly, judging by the stability of the coefficients in the two variants of the model, we conclude that the COVID-19 pandemic did not change the character of the relationship between push factors and quit considerations, but it elevated the prevalence of phenomena such as mental health and work-related distress, which in return led to more quits. 

\section{Discussion}\label{sec:discussion}
The COVID-19 pandemic shook the global labor market like no other economic recession that we have on record. It led to the Great Resignation in 2021, a record high quit rate in the U.S. and considerably high quit rates in other countries. While traditional economic forces, such as high labor shortages and the resulting wage increases played an important role, the media is now often citing burnout, toxic work environment and desperation as leading causes. Such motivations cannot be easily studied using official survey and administrative data, and one can remain sceptical of bombastic media headlines that use private sources of data or small-scale surveys. In this paper we take a different approach and use data from Reddit. We focus on `r/jobs', where more than a hundred thousand people shared their work-related questions and concerns, discussing issues that emerge more organically and are more intimate than what official statistics can cover. Through studying the evolution of the work discourse using sentiment and text analysis, we shed light into how the pandemic affected workers and find some evidence that, along with some of the usual causes of rise in quits in recovery periods such as job switching, mental health concerns may have been one of the drivers of the Great Resignation. Here we summarize our main findings and discuss their implications. 

Our first finding, which also serves as validation, is that the evolution of the `r/jobs' discourse resembles the dynamics of U.S. labor market. We show this first by comparing the shares of quit- and fired- related posts with the U.S. quit and layoff rates, and then by showing that after the pandemic topics related to remote work sharply increased their prevalence, while the topic about commuting declined. These findings contribute to the literature using social media data to measure, and in some cases predict, changes in the economy \citep{antenucci2014using,bollen2011twitter}. Our study supports the use of Reddit as a real-time socio-economic observatory where one can study work and labor when administrative and  official survey data is not available or to complement it. 

The sentiment analysis revealed that the overall sentiment became more negative during the first quarter of the pandemic. This finding is inline with the literature documenting that the pandemic had a negative psychological impact \citep{ashokkumar2021social,xiong2020impact}. We find that the overall sentiment returned to pre-pandemic levels in 2021. This could be due to the better labor market conditions or adaptation of workers to new working conditions (some studies have also shown psychological numbing in sentiment after the pandemic \citep{dyer2020public}). Using a difference-in-differences analysis we find that when people talked about quitting, as compared to talking about work more generally, they were significantly more negative during the first quarter of the pandemic. Anger, fear, sadness, and disgust all peaked among quit-related posts at the onset of the pandemic. This portrays the very difficult situation that some working people were in and contributes to the literature documenting the general \citep{kwon2022understanding} and career specific \citep{akkermans2020covid} psychological effects of the pandemic.

Using topic modelling, we document how the overall work discourse changed since the pandemic started. The topics that increased their prevalence the most include switching jobs, mental health, work-related distress, and remote work topics. In contrast, commuting was one of the topics that decreased its prevalence the most. Besides serving as validation, these results contribute to a more complex understanding of workers. We identify factors other than wages, such as mental health concerns, that may influence workers decisions. In this way, we contribute to a growing literature studying meaning of work \citep{rosso2010meaning,nikolova2020makes}, and can be a starting point to help guide models that go beyond modelling workers as wage-leisure maximizers and portray human action in a more realistic manner. 

Our most important finding is that mental health and work-related distress likely contributed to the Great Resignation. Two results support this conclusion: First, mental health and work-related distress significantly increased their prevalence since the start of the pandemic, in quit posts relative to nonquit posts. Second, our multiple regression model revealed that mental health concerns and work-related distress topics are related to a higher probability of a post being about quitting. We also find some evidence that some relief came to mental health concerns and work distress in the last two quarters of the Great Resignation, when the prevalence of \textit{mental health} and \textit{work-related distress} decreased. One possible reason for this relief may be the better labor market conditions of the year 2021. These findings contribute to the understanding of the causes of the Great Resignation. Furthermore, we also contribute to the management literature by providing empirical evidence to the theoretical notion that shocks at the societal level can prompt turnover cognitions \citep{morgeson2015event}.

Our last finding is that the relationship between mental health concerns and quitting did not change with the pandemic. In other words, people that suffered from mental health issued were not more likely to quit after the pandemic. Instead, it was the increase in prevalence of mental health concerns that contributed to the rise in quits. This result contributes to the literature documenting the effects the pandemic had on people's mental health \citep{xiong2020impact}.

Our research is not without its limitations. First, our results are based on the user population of `r/jobs', which is not representative of neither the world nor the U.S. labor force. Furthermore, it is difficult to extract the demographic characteristics and employment status of users. This is a limitation since the pandemic had heterogeneous affects among people of different gender, age, and occupation \citep{adams2020inequality,cook2021who,del2020supply}. Second, we cannot guarantee that the quit-related posts translate into an actual quit -- the fact that a user talks about quitting does not guarantee the person will indeed quit their job. Furthermore, if there is a quit, we do not know the precise timing. People may discuss a past quitting experience or planning to quit. Third, our results rely on sentiment analysis and topics modelling, methods that have limitations \citep{iliev2015automated}. For example, topic modelling uses a bag-of-words approach that in our study leads to multi-topics, where different topics get bundled together due to the same words being used in different contexts. 

Nonetheless, we have made strong efforts to reduce and understand the effects of these limitations. We have leveraged on  \cite{waller2021quantifying} study to uncover the characteristics of `r/jobs' users. Furthermore, we are one of the few studies \citep{von2021behavioral,seraj2021language} using Reddit that go through several hundred of posts, follow authors posting history, and manually encode self-disclosed information to understand the composition of the sample under study. Our results show that our sample has more women and young workers than the U.S. working population. Given the strong challenges working women faced during the pandemic \citep{adams2020inequality} and studies suggesting that the Great Resignation was driven by young adults~\citep{cook2021who}, we consider our sample population to be of interest for the topic. Although we cannot guarantee that people who talk about quitting will actually quit, previous studies have shown that cognitions about quitting are a strong predictor of quit behavior \citep{rubenstein2018surveying}. To overcome topic modelling limitations, we used a Structural Topic Model, which is state-of-the-art in social science studies \citep{hannigan2019topic}. We consider that, despite the mentioned limitations, the advantages of using Reddit data and Natural Language Processing methodologies out-weight the caveats. Digital trace data is available in real time and allows for more personal and in-depth expression than surveys. In this sense our work does not substitute but complements traditional studies based on surveys or economic models.

Finally, this work can help guide some policy and business strategies. Our results suggest that some of the distress caused by the pandemic \citep{gruber2021mental} are linked to working conditions. 
This underscores the importance of designing work policies to reach the 2030 mental health targets of the  World Health Organization \citep{world2021report}. Furthermore, our work suggests that businesses trying to retain their workers or hire new people should consider prioritizing the mental health of their workers through, for example, the (re-)design of jobs \citep{harvey2017can} or by providing company-sponsored therapy sessions for employees \citep{joyce2016workplace}.

\section{Materials and Methods}\label{sec:methods}
\subsection{Data}

\paragraph{Extraction and filtering of reddit posts}

We extract Reddit submissions posted in the subreddits `r/antiwork', and `r/jobs' between January 2016 and December 2021 (inclusive) using Pushshift API \citep{baumgartner2020pushshift}. For each post, we have, among other data, the username of the author, the date when it was posted, title, text content of the post, and flair. 
We remove all posts that were flaired as spam, scams, removals or moderator or bot posts as well as those with no title and text content. Furthermore, we focus only on years 2018-2021, where the majority of the posts were made. This leaves 134,212 `r/antiwork' posts and 269,647 `r/jobs' posts.

Since we focus our analysis on `r/jobs' we do some additional cleaning for this subreddit. First, we find that there are a small number of authors that have many posts on the subreddit (e.g., there were roughly 70 authors who had more than 50 posts). To avoid biasing our study towards authors that post a lot but may not represent an average user, we remove all posts that correspond to authors who have more than 10 posts in the subreddit. This leads to roughly 190,000 posts to analyse in `r/jobs'.

\paragraph{Text pre-processing}
We first concatenate the title and body of the post, referred from here onwards as the text.  We lowercase all text and remove special characters and sequences such as quotes, newlines, parenthesis, etc. We unravel the acronyms ``pto" and ``wfh" to ``paid time off" and ``work from home''; we remove the acronym ``tldr" (too long didn't read). We homogenize different ways to refer to the COVID-19 pandemic to `covid', e.g., we replace `coronavirus' and `covid-19' with `covid' (we do not replace `pandemic'). For handling negation, as our method is dictionary-based, we create ngrams following negation words using part-of-speech tagging, i.e., if a negation word is followed by a noun, an adjective, or a verb within a three-word window, an underscore will be added between the corresponding words to be ignored by the emotion lexicon. For example, the phrase 'not happy' will become 'not\textunderscore happy'. This way, 'happy' is not classified as a positive word as it would be the case without negation handling. 

For topic modelling, we take the following additional steps. We remove a set of common stopwords (i.e., non-content words such as ``the", ``do", or ``throughout") using Scikit-learn's \citep{pedregosa2011scikit} list of English stopwords to which we add ``ai", ``im", ``m", ``s", ``ve", ``w", ``d", ``ive", ``id", ``itll". Then, we form bigrams or trigrams from ordered set of words that commonly appear together, and lemmatize unigrams. After n-gram formation, we delete tokens which are overly common (``job", ``like", ``just"), boilerplate (``andor", ``http", ``amp"), or idiosyncratic to Reddit (``long\_post", ``sorry\_long", ``rjobs", ``hey\_guy"). We then create the final document-feature-matrix (dfm) restricting the used terms to those appearing in a minimum of .5\% and a maximum of 99\% of all texts.

Our pre-processing workflow follows the recommendations by \citealp{Hickman_preprocess_2022} for both closed and open vocabulary approaches.

\paragraph{Posts labels and control group}
Our unit of analysis are time stamped Reddit posts from 2018-2021, where a post is an original submission without follow-up comments. We aggregate time stamps into monthly and quarterly intervals to have a significant amount of posts for each period. We use keywords to label posts as quit- and nonquit- related. To do this, we go through the text of a post and search for quit-related keywords. If we find at least one keyword related to quitting in the post, we label said post as quit-related, otherwise, we label it as nonquit-related. We also use keywords to label fired-related posts. Below we explain in more detail how we use keywords. 

We distinguish between quit and nonquit (control) posts using keywords related to quitting. We use the following keywords: \textit{resign*, quit*, leav* (my, the, a) job, left (my, the, a) job, bow* out, (week*, my, a, the) notice, switch* (*) job, chang* (*) job, look* for (*) new job}, where * denotes any character and () denotes optional character sequence. 

\footnote{%

We fortified the keywords to match any number of whitespace between the words. In the text, * represents any number of characters allowing for different forms of words, e.g., \textit{resign*} captures \textit{resignation, resigned, resign} and some typos, e.g., \textit{resigning}. For some keywords, we allow for optional words in between keywords (*), e.g., \textit{chang* (*) job} matches both \textit{change a job} and \textit{change job}. 
} 

We did not rely on posts that were tagged with the flair \textit{Leaving a job} since we are aware that quit-related posts could have been flagged with other flairs such as \textit{Job offers} and \textit{Job searching}, in the case of switching or looking to switch jobs, or \textit{Unemployment} in the case of quitting a job without another one lined up. 

The purpose of the control group is to help discern how the content of quit-related posts has changed since the onset of the pandemic not only relative to the pre-pandemic levels, but also relative to posts that are not quit-related. To do this, we build a control group of nonquit-related posts. Initially, we considered building the control group from a subsample of other subreddits. However, this method appeared to yield a poor control group, due to the dominance of topics that have little to do with people's work, and, likely, due to the different demographic structure of the contributors. Hence, we build the control group using `r/jobs' posts as follows. First, we exclude all quit-related posts and all posts made by authors that had quit-related posts. From this selected group of nonquit-related posts, we take a random sample of $m_y$, where $m_y$ is the number of quit-related posts for year $y$ (4996, 6860, 6080, and 8080 for 2018 to 2021 respectively). Therefore, in each year, the control group has the same size as the treatment group. Additionally, for the sentiment analysis we drop a small fraction of posts from the treatment and control group that have less than 15 words. The size of the treatment and control group remains roughly of the same size each year. 

To do the multiple regression analysis, we did not fit the Structural Topic Model to the whole sample of `r/jobs' posts due to computational constraints. Instead, we required a sample of quit- and nonquit- related posts with the same composition of quit- and nonquit- related posts as the general `r/jobs' population of posts. To get a sample with the sample composition of `r/jobs' we did the following. We started from the sample of posts that was half quit- and half nonquit- related and selected all the nonquit-related posts. Then, for each quarter, we randomly selected quit-related posts with a probability equal to the `r/jobs' proportion of quit- and nonquit-related posts. This process led to a subsample of roughly 29,000 posts.

\subsection{Structural Topic Modelling}

Topic modeling is an unsupervised clustering method aimed to discover latent topics from texts automatically \citep{blei2012probabilistic}. The basic assumption is that words which co-occur in documents (Reddit posts in our case) discuss the same subject. In the version of topic modeling in which topics are allowed to correlate, documents represent a mixture of topics, and the final model computes the prevalence of each topic in a per-document basis. STM allows for topic correlations, i.e., to account for the fact that some combinations of topics are more likely to co-occur. Furthermore, STM allows  within-document topic prevalence to differ across metadata such as the time in which the text was written. STM has been used to understand temporal sense-making processes following terrorist attacks \citep{fischer2019collective}, gender and race (among others) differences in the challenges faced by leaders \citep{tonidandel2021using}, and the impact of corporate funding in the production of polarizing climate-change related texts \citep{farrell2016corporate}. 

We fit STMs using the packages \textit{stm} \citep{roberts2019stm} and \textit{quanteda} \citep{benoit2018quanteda} in the statistical environment \textit{R} \citep{r2020r}. Specifically, we fit STMs with the following simple difference-in-differences equation:
\begin{equation}
y_{i} = \alpha + \beta_1T_{t} + \beta_2Q_{i}+ \beta_3T_{t}Q_{i} + \epsilon_{i}
    \label{eq:DiD_simple}
\end{equation}
Here, the prevalence of a topic $y$ in a post $i$ is determined by $T_t$, a dummy capturing if the post was published before or since the onset of the pandemic (March 1, 2020), $Q_i$, the dummy variable for quit- (vs. nonquit-) related posts, the interaction of both variables $T_tQ_i$ and the error term $\epsilon_i$. We fit STMs with equation (\ref{eq:DiD_simple}) at K [K= 5 - 200] incrementing the number of topics by 5. We then plot the average exclusivity and semantic coherence \citep{mimno2011optimizing, roberts2014structural} of each model to choose the optimal K following the approach by \cite{hofstra2020diversity}. The semantic coherence, measured by the frequency in which the words in a topic co-occur, reflects the internal consistency of a topic. Maximizing semantic coherence, however, is not enough to secure a good solution, as less topics will mechanically contain more co-occurring words. Thus, researchers recommend to look at the combination of semantic coherence and exclusivity \citep{roberts2014structural, hofstra2020diversity}. Exclusivity measures the extent to which the words in a topic differentiate from those of others. There is a trade-off between both measures. Topic coherence is positively influenced by the probability of the words it contains; the higher the probability of the words, the more likely they will co-occur. However, the same words will lower the exclusivity of a topic because of their inclusion in multiple topics. 

\begin{figure}[h]
    \centering
\includegraphics[width = 0.95\textwidth]{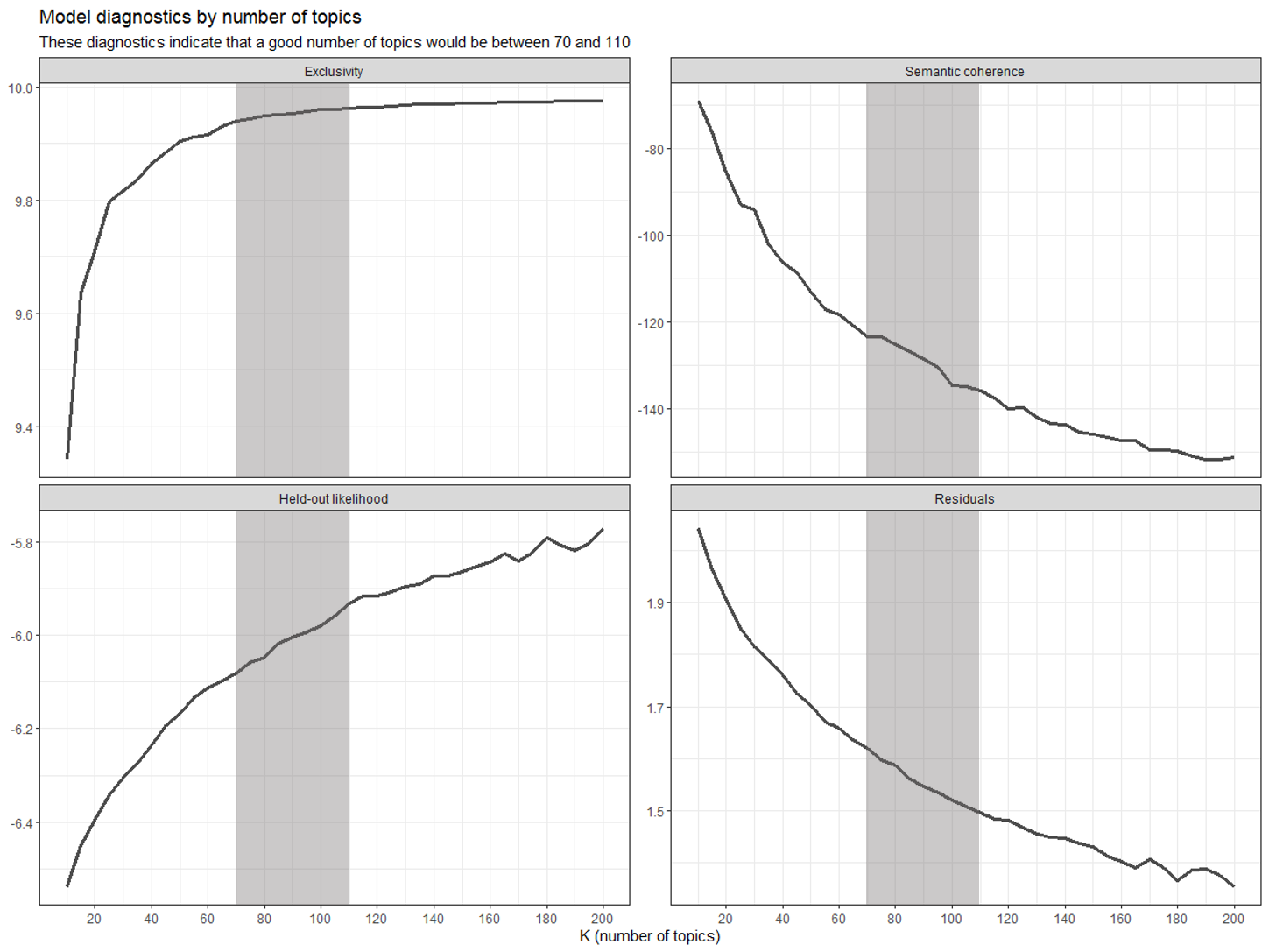}
    \caption{Diagnostics of models including from 5 to 200 topics, augmenting by 5. The shaded area represents the values of K we consider for the fit. In the Supplementary Material~\ref{sec:SI_topicmodelling} we run robustness checks with the floor and ceiling of this area.}
    \label{fig:STM_searchK}
\end{figure}

Following \cite{hofstra2020diversity}, we choose the optimal K by locating the point in the graph in which exclusivity and semantic coherence plateau, such that incrementing the number of topics does not yield higher exclusivity or lower coherence.  The results are visible in  Figure~\ref{fig:STM_searchK}, which also includes the held-out likelihood (the fit of each solution in a hold‐out sample) and the residuals of the different models. We observe that the exclusivity augments steeply until around 50 topics, when adding more topics slowly leads to lower increases in exclusivity. We choose 90 topics because this seems to be the plateau in the graph. 

Once we fit the model with 90 topics we label the topics  following the recommendations by \cite{debortoli2016text}. Specifically, two members of the research team independently label each topic guided by the top 10 terms chosen using the frequency-exclusivity scoring ([FREX]; \cite{roberts2013structural}) and through careful reading a sample of 25 documents associated with each topic selected using the \textit{findThoughts} function of the \textit{stm} package. A third member reads the designations resulting from this task and decides the final label. Disagreements are solved by discussion. 

We distinguish between three types of topics based on their inner consistency - clear topics (CL), multi-topics (MT), and boiler plate topics (BT). Clear topics, as the name suggests, are readily interpretable topics. Multi-topics are topics that include more than one clearly interpretable topic. Boiler-plate topics tend to be centered around certain words but do not have an inner consistency, e.g., a topic consisting of posts that contain the word `look'. While boiler-plate topics provide little intrinsic meaning, they are helpful to encapsulate noise and avoid spillovers to other more meaningful topics \citep{dimaggio2013exploiting}. For reference, we label boiler-plate topics with the word they seem to cluster around, but discard these topics from our analysis.  In total, we identified 68 clear topics, 10 multi-topics, and 12 boiler-plate topics. 

\section{Acknowledgements}
We would like to thank Ken Benoit, Theresa Gessler, Frank Neffke, Max Pellert, Taha Yasseri and the other participants of the Complexity Science Hub Workshop on the Great Resignation for their feedback and discussions. We are also thankful to Mikołaj Biesaga for providing the code in Python to collect Reddit data and to Sebastian von Beck and Fariba Dorpoush for their work as Research Assistants. MDRC acknowledges funding from James S. McDonnel Foundation. MSF acknowledges funding from Science Foundation Ireland under Grant number 18/CRT/6049. We would also like to thank the organisers of SICSS London 2021 for helping us kick-start this project and the Complexity Science Hub for funding the Workshop on the Great Resignation. Finally, we would like to thank Joshua Becker, Marco Pangallo, and Ingo Weller for their feedback on the manuscript.

\FloatBarrier
\bibliographystyle{agsm}
\bibliography{references}
\FloatBarrier

\newpage
\appendix
\renewcommand{\thesection}{S~\arabic{section}}    
\renewcommand{\thefigure}{S~\arabic{figure}}
\setcounter{figure}{0}
\renewcommand{\thetable}{S~\arabic{table}}
\setcounter{table}{0}

\section*{Supplementary Material for\\ Mental health concerns prelude the Great Resignation: Evidence from Social Media
}
\author{R. Maria del Rio-Chanona$^{1,\dagger}$, Alejandro Hermida-Carrillo$^{2,\dagger}$,  \\Melody Sepahpour-Fard$^{3,4}$, Luning Sun$^{5}$,   Renata Topinkova$^{6,7}$, and Ljubica Nedelkoska$^{1,8}$\\
\\
\footnotesize{$^{1}$ Complexity Science Hub, Vienna}\\
\footnotesize{$^{2}$ LMU Munich School of Management}\\
\footnotesize{$^{3}$ Science Foundation Ireland Centre for Research Training in Foundations of Data Science}\\
\footnotesize{$^{4}$ Department of Mathematics and Statistics (MACSI), University of Limerick}\\
\footnotesize{$^{5}$ The Psychometrics Centre, University of Cambridge}\\
\footnotesize{$^{6}$ Institute of Sociology of the Czech Academy of Sciences}\\
\footnotesize{$^{7}$ Faculty of Arts, Charles University}\\
\footnotesize{$^{8}$ Harvard Growth Lab}\\
\footnotesize{$^\dagger$ These authors contributed equally}\\
}


\section{The U.S. labor market before and during the pandemic}\label{sec:SI_economics}
In this section we discuss in more detail U.S. economic situation relevant for the Great Resignation. In particular we discuss factors that stand out in the U.S. economy as plausible drivers of the Great Resignation. 

\begin{figure}[h]
    \centering
\includegraphics[width = .49\textwidth]{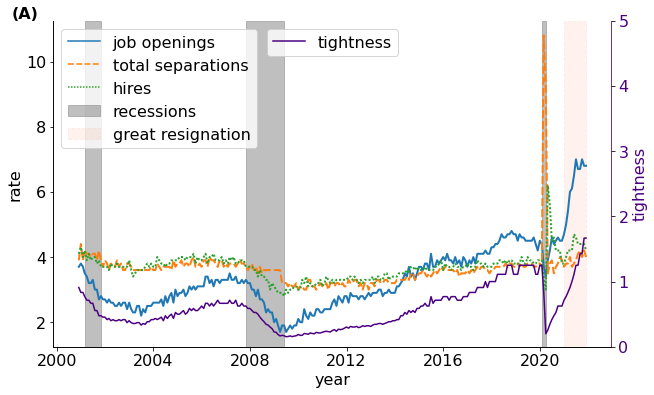}
 \includegraphics[width = .4\textwidth]{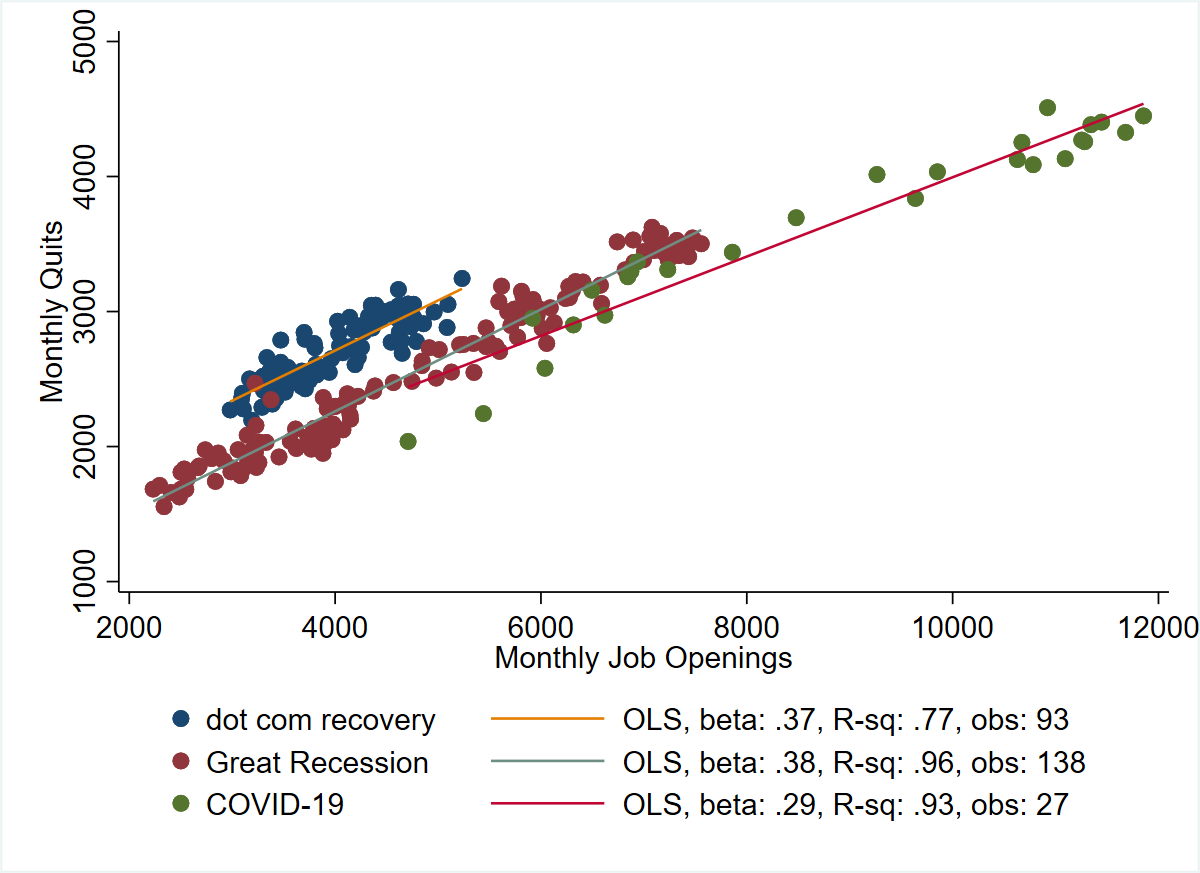}
    \caption{\textbf{US labor market} Left. U.S. labor market statistics from December 2000 to December 2021. Recession periods are marked with shaded grey areas. The dashed line corresponds to March 2020 and the colored area to the Great Resignation period (year 2021). Right. The relationship between quits and job openings in recovery periods. }
        \label{fig:apx_labor_market_summary}
\end{figure}

One factor we discussed in the main text is the pro-cyclical behaviour of the quit rate, job openings and labor market tightness (i.e., the ratio between the number of job openings and unemployed workers). In addition to the quit rate shown in the main text, here in Figure~\ref{fig:apx_labor_market_summary} (left) we show the pro-cyclicality of the job openings and labor market tightness of the U.S. economy. Although the pro-cyclicality likely played a role in the Great Resignation, none of the recession aftermaths since the 1990 created a similar surge in quits \citep{BLS2022JOLTS_1, davis2014labor}. During 2021, total quits and quits as a share of total separations reached a new record. Between 2001-2020 54\% of all job separations were quits, in contrast, quits accounted for 69\% of all job separations in 2021. In Figure ~\ref{fig:apx_labor_market_summary} (right) we show the relationship between the quit rate and job openings during recovery periods. We find that there has been a weakening between the quit and the hiring rate: in the twenty years prior to the pandemic, one additional job opening was associated with 0.38 additional job quits during recovery periods. In the COVID-19 recession recovery, an additional job opening has been associated with 0.29 quits. 

Instead of switching jobs, more people than ever seem to be opting into self-employment and start-ups \citep{wsj2021}. Furthermore, some of the registered quits may also be quits that people delayed or postponed during the pandemic \citep{cook2021who}. Another pull factor in the U.S. that in theory could have motivated people to quit was the stimulus policy enacted during the pandemic. However, \cite{petrosky2020reservation} and \cite{boar2020dynamic} show that these checks were too small and too temporary to make it worthwhile for people to quit in large numbers. 

In the main text we discussed the push factors the pandemic unleashed and how these may have driven people to quit. Some of the effects of these push factors are recorded in moves to non-participation among older workers \citep{heggeness2021telework, BLS2022_1}. After all, elderly workers were at disproportionate risk of suffering from COVID's health impacts. However, young people were also affected by the push factors through difficulties adapting to home office, struggles with daycare and school closures and burnout. As we show in Appendix \ref{sec:SI_reddit_population}, our sample of study is mostly young adults, therefore we most likely capture the effects of push factors not included in early retirement. 

\FloatBarrier

\section{Reddit and work} \label{sec:SI_rjobs}
In this section we present additional information and statistics of the overall population of Reddit and discuss the subreddits `r/jobs' and `antiwork'. For `r/jobs', the subreddit we focus our analysis on, we provide a detailed analysis at the end of the section. 

\subsection{The characteristics of the Reddit population} \label{sec:SI_reddit_population}

Reddit, self-proclaimed ``The front page of the internet", is an online platform ranked among the ten most visited websites in the world \citep{alexa}. Reddit is open worldwide, but most users reside in English-speaking countries; the largest share of traffic (roughly between 40 - 50\%) comes from the U.S., followed by the U.K. at 7.9–8.2\% and Canada at 5.2–7.8\%  \citep{alexa, similarweb}. According to Reddit Inc., U.S.-based adult users are mostly young adults: $58\%$ are between 18-34 years old and $42\%$ are 35 or older \citep{RedditInc}. Gender is slightly skewed towards males, with 56\% of users being male \citep{RedditInc}. In contrast, the U.S. 2021 labor force was distributed roughly equally between the age brackets 18-34, 35-49, and above 50 ($34\%$, $31\%$, $33\%$ respectively, with the reminder $2\%$ corresponding to 16-17-year-olds) and males accounted for $53\%$ of the U.S. labor force\footnote{
We calculated the labor force demographic distribution from BLS data \url{https://www.bls.gov/cps/lfcharacteristics.htm\#laborforce}
}. 

While the aggregated statistics described above provide a rough overview of the general population and the difference with the U.S. labor force generally, there is large heterogeneity among the population of each different subreddit depending on the central topic being discussed \citep{waller2021quantifying}. There are a number of different subreddits related to work and jobs including `r/work', `r/jobs', `r/careerchange', and r/antiwork. Since `r/antiwork' and `r/jobs' have the highest number of users and 'r/antiwork' received considerable attention on the topic of the Great Resignation in the media  \citep{Rogers_2022}, we considered both subreddits for our study. 

\subsection{`r/jobs' vs. `r/antiwork'} \label{sec:SI_jobs_vs_antiwork}

\begin{figure}[h!]
    \centering
\includegraphics[width = 0.75\textwidth]{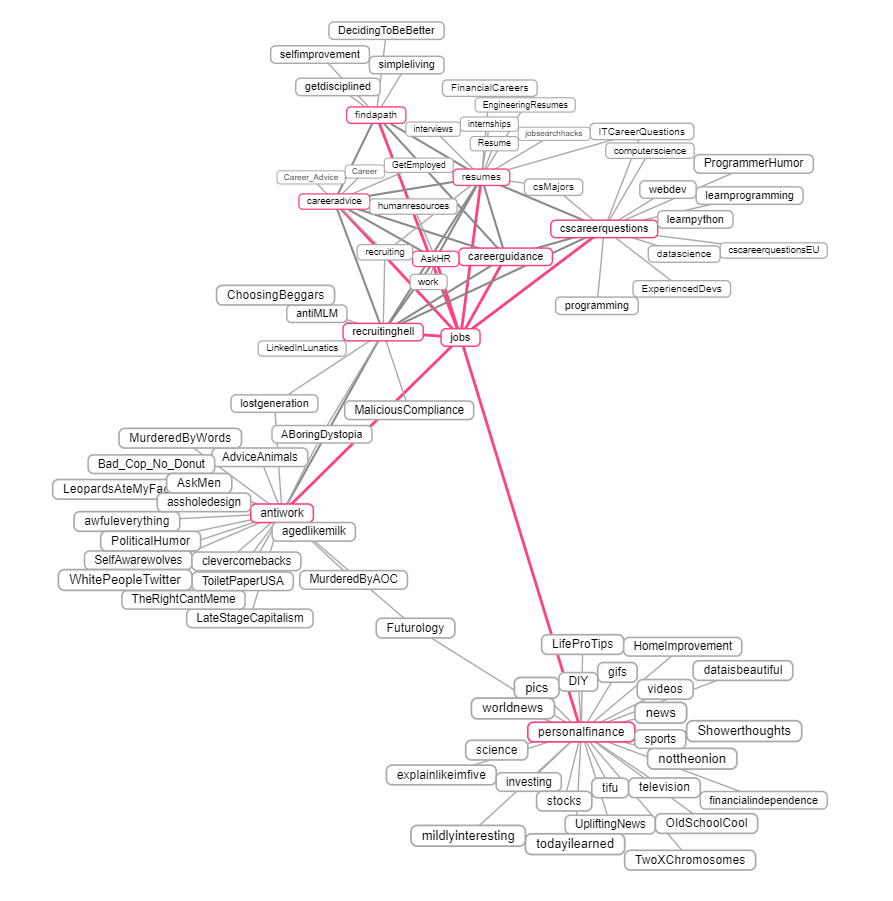}
    \caption{\textbf{Co-occurrence of subreddit users} In this network, nodes represent subreddits, and ties denote that users who commented to subreddit A, also commented to subreddit B.}
    \label{fig:sayit}
\end{figure}

To choose a subreddit suited for our study from 'r/jobs' and 'r/antiwork', we looked into i) the interests and ii) the posting activity across time of the users of both subreddits. We explored the useres' interests using the online tool ``sayit'' (available online at \url{https://github.com/anvaka/sayit}), which computes Jaccard similarity between subreddits based on the subreddits where users comment.  As shown in Figure \ref{fig:sayit}, although there is some overlap in the interest of users of these two subreddits, there are considerable differences. `r/antiwork' is strongly political, and its users often post to other left-leaning (including openly ``anticapitalist") political subreddits, such as `r/LateStageCapitalism', `r/ABoringDystopia', or `r/MurderedbyAOC'. In contrast, users posting to `r/jobs' often post to other career-oriented subreddits, e.g., `r/AskHR', `r/careerguidance', or `r/resumes'. To examine the posting activity across time, we extracted all posts of both subreddits from 2016 to 2021. Table~\ref{tab:subreddits} shows the raw number of posts (i.e., without filtering out spam and moderator posts) in each subreddit before and after the onset of the COVID-19 pandemic. Although `r/antiwork' has more members than `r/jobs', its popularity soared after the onset of the pandemic. In fact, less than 10\% of the posts are before the pandemic. Instead, `r/jobs' has more equal distribution of posts before and after March 2020. 

\begin{table}[h]
\centering
\begin{tabular}{c|cc}
Measure & `r/antiwork' &  `r/jobs' \\ \hline
Members (as of February 4th 2022) & 1.5 million   & 0.65 million       \\
Total number of posts (2018 - 2021) & 134,212    & 269,647       \\
Number of pre-pandemic posts (2018 - March 2020)   & 10,962    &      142,514    \\
\end{tabular}
\caption{\textbf{Overview of work-related subreddits} Number of submissions from January 1st, 2018 to December 31st, 2021. These are the raw numbers of overall submissions and include some spam and moderator posts. For later analysis we filter several of these (see \emph{Methods} section for details).
}
\label{tab:subreddits}
\end{table}

Since `r/jobs' has a more balanced distribution of posts across time, and because its user composition is less politically skewed, we conclude that `r/jobs' is a better subreddit to study the Great Resignation and continue to understand its population in the next subsection. 

\subsection{The population of `r/jobs'} \label{sec:SI_jobs_population}

We examine the `r/jobs' population in two ways. First, we use the results in the Supplementary Material of a recent study by \cite{waller2021quantifying}, who used neural-embeddings to characterize the populations of various of the most popular subreddits in terms of age, gender and U.S. political partisanship. In particular, each of the subreddit has an age coordinate between -0.61 and 0.62, where a more negative value indicates a younger population relative to the general Reddit population; and a gender coordinate between -0.35 and 0.52, where a more negative value indicates a more male population relative to the general Reddit population. The `r/jobs' age coordinate is 0.21 and female coordinate 0.11. In other words, in comparison to the average subreddit, `r/jobs' has an older population and a slightly higher female representation.

 Second, we examine the posting history of a subsample of `r/jobs' users to infer their demographic characteristics using an approach similar to \cite{von2021behavioral}. Specifically, we randomly select the authors of 200 posts before and 200 posts after March 1st 2020, i.e., the date we use as the onset of the pandemic, from our `r/jobs' post sample. With the help of a research assistant, we hand-code self-disclosed demographics in the Reddit posting history of each author (e.g., a user writing ``f/20" in a post). We are able to record self-disclosed gender for ca. 40\%, age for ca. 29\%, education level for ca. 38\%, and country of residence for ca. 32\% of all 400 users. We present the results in Table \ref{tab:rjobs_sample_composition}.

\begin{table}[h]
\centering
\begin{tabular}{p{7cm} c |c}
  &  Pre 01.03.20 & Post 01.03.20\\ \hline
Gender$^a$  \\ \hline
Female  & 51\% & 68\% \\ \hline
Age$^b$ \\ \hline
17 years old or younger & 10\%  & 6\%  \\ 
between 18-34 years & 78\%  & 92\%  \\ 
between 35-49 years & 12\%  & 2\%  \\  \hline
Education$^c$  \\  \hline
High school degree or less & 14\%  & 9\%  \\ 
Undergraduate degree & 75\% & 76\%  \\
Graduate degree & 11\%  & 15\% \\ \hline
Country of residence$^d$ \\  \hline
U.S. &  81\% & 70\%  \\ 
Canada &  6\% & 8\%  \\
U.K  &  3\% & 13\%  \\
Elsewhere &  10\% & 10\% \\
\end{tabular}
\caption{\textbf{Demographics of users of r/jobs.}  Reported are aggregates of self-disclosed characteristics found within the posting history of a random sample of 400 users (200 pre and 200 post the onset of the COVID-19 pandemic). Undergraduate and graduate degree statistics include both individuals reporting having completed as well as those reporting dropping out or currently studying a degree.\newline
 Characteristics for $^a$ were identified on 158 (89 pre and 69 post 01.03.2020) authors, for $^b$ on 115 (68 pre and 47 post), for $^c$ on 153 (88 pre and 65 post), and for $^d$ on 127 (64 pre and 63 post).
}

\label{tab:rjobs_sample_composition}
\end{table}

These results complement those of \cite{waller2021quantifying} and, taken together, suggest that we are studying the Great Resignation through a population of mostly U.S.-based working age young adults. After the onset of the pandemic, the user population became more international, although the U.S. remained by far the most common country. Interestingly, after the onset of the pandemic, the female population became the majority. This finding may be caused by the disproportional burden women faced during the pandemic at work \citep{adams2020inequality}, leading them to rely on community support to help them overcome such issues.

When interpreting these results, we must take into account that we are relying on self-disclosed information and that the probability that a user mentions their gender or country of residence may differ for different groups of users or might have changed over time. For example, during the pandemic users might have been more likely to report they were not from the U.S. to provide context to other redditors. Therefore, we provide the results in Table \ref{tab:rjobs_sample_composition} to provide context, but we do not consider them reliable enough to incorporate them into our statistical analyses.  

\subsection{The dynamics of `r/jobs'}
\begin{figure}[h]
    \centering
\includegraphics[width = .45\textwidth]{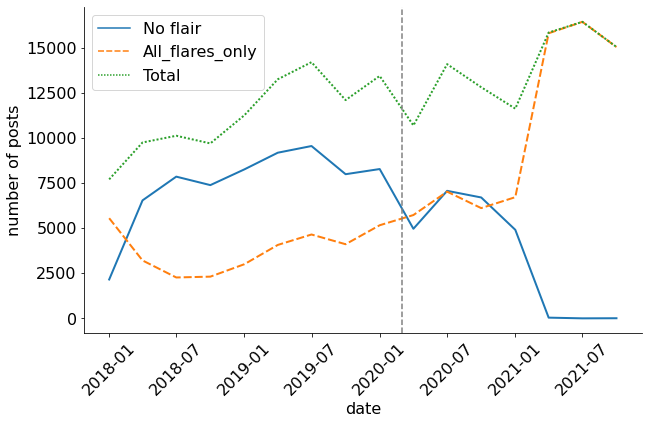}\includegraphics[width = .45\textwidth]{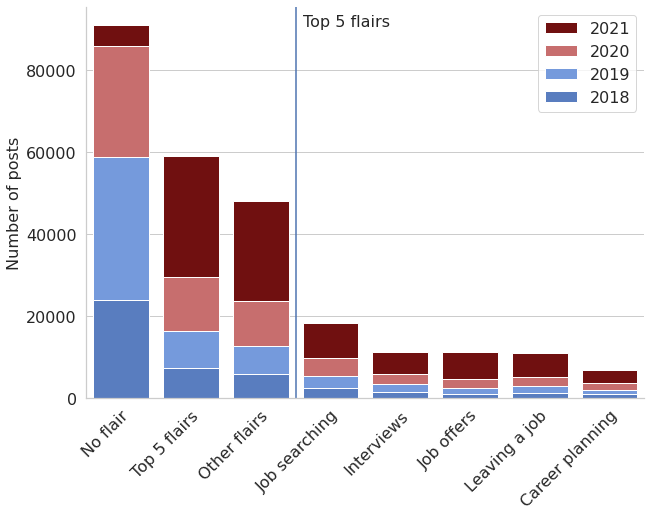}
    \caption{\textbf{Distribution of r/jobs posts across time and flairs.} Figures show the total number of `r/jobs' posts after filtering. Left: Total number of posts with and without flairs (2018-2021). Right: Total number of posts with no flair, top 5 flairs, and remaining flairs. Blue represents pre-pandemic years (2018 and 2019), red represents pandemic years (2020 and 2021).  }
        \label{fig:flares_noflares_time_bar}
\end{figure}

In this subsection, we provide additional information on the `r/jobs' subreddit. Table~\ref{tab:rjobs_nposts} shows the total number of posts from January 2018 and December 2021 in `r/jobs' after filtering out spam, moderator posts, advertisements, etc. (see \emph{Methods} section for details). Figure~\ref{fig:flares_noflares_time_bar} (left) shows the number of posts across years split by post with a flair, those without, and both. This figure shows a steady increase in the total number of posts in `r/jobs'. The number of posts has almost doubled between 2018 and 2021. 

\begin{table}[h]
\centering
\begin{tabular}{c|c}
Measure (after filtering)  &  `r/jobs' \\ \hline
Number of posts (2018-2021) & 198081        \\
Number of posts pre-Covid (2018- March 2020)  & 88092      \\
\end{tabular}
\caption{\textbf{Number of posts in r/jobs.} These figures show the total number of `r/jobs' after filtering.
}
\label{tab:rjobs_nposts}
\end{table}

Reddit allows users to add a tag to their posts to specify the content. For example, `r/baking' allows \textit{Recipe} and \textit{No recipe} flairs to signal whether posts contain a recipe or not. Flairs are pre-defined, specific to each subreddit, and mutually exclusive, i.e., users select flairs from a limited number of available flairs, cannot create new ones, and can only use one flair per post. In our sample of `r/jobs' posts there are 36 flairs (excluding those related to spam, moderation, etc.). The five most popular flairs (\textit{Job searching}, \textit{Interviews}, \textit{Job offers}, \textit{Leaving a job} and \textit{Career planning}) account for roughly 50\% of the flaired posts.

The bar plot in Figure~\ref{fig:flares_noflares_time_bar} (Right) shows the number of posts by year without a flair, with one of the top five flairs, or with one of the other flairs. This figure also shows the number of posts in each of the top five flairs, the most popular flair being \textit{Job searching}. The name of the rest of the flairs are listed below:

\begin{multicols}{4}
\begin{enumerate}
    \item `Applications'
  \item `Article'
  \item `Background check'
  \item `Career planning'
  \item `'Companies'
  \item `Compensation'
  \item `Contract work'
  \item `Covid-19'
  \item `Discipline'
 \item  `Education'
  \item `Evaluations'
  \item `HR'
  \item `Internships'
  \item `Interviews'
  \item `Job offers'
  \item `Job searching'
  \item `Layoffs'
  \item `Leaving a job'
 \item  `Networking'
  \item `Office relations'
  \item `Onboarding'
  \item `Post-interview'
  \item `Promotions'
  \item `Qualifications'
  \item `Recruiters'
  \item `References'
  \item `Rejections'
 \item  `Resumes/CVs'
  \item `Startups'
  \item `Temp work'
  \item `Training'
  \item `Unemployment'
  \item `Work/Life balance'
\end{enumerate}
\end{multicols}

\begin{figure}[!h]
    \centering
\includegraphics[width = .7\textwidth]{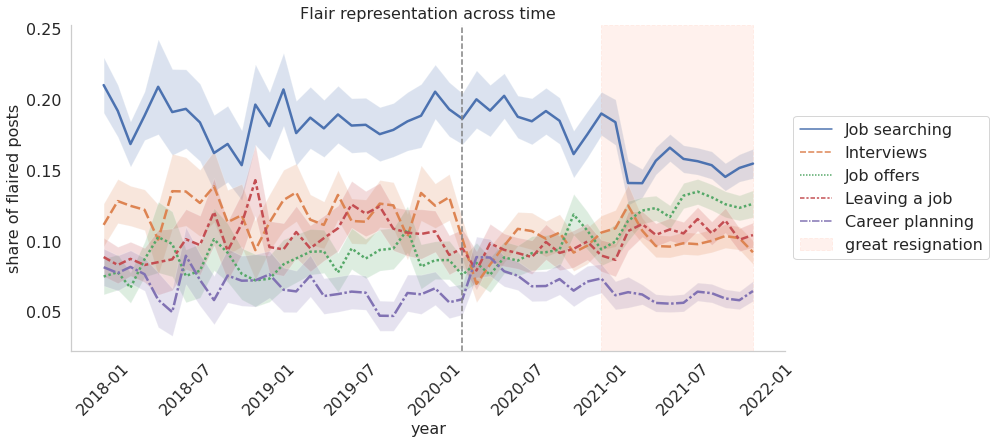}
    \caption{\textbf{r/jobs posts across time.} This figure shows the share of posts tagged for each of the five most popular flairs. Approximately $50\%$ of all posts are tagged with one of the top five flairs.}
        \label{fig:flares_through_time}
\end{figure}

To further validate whether `r/jobs' reflects what happened in the U.S. labor market, we analyze how the popularity of different `r/job' flairs developed over time. Since the five most popular flairs account for roughly 50\% of the flaired posts we focus the overview  on these flairs. As Figure ~\ref{fig:flares_through_time} shows, the popularity of the top five flairs (i.e. the share of posts corresponding to each flair) remained roughly constant until April 2021. In this month, which is also the month when the quit rate reached a record high in the U.S., there were some abrupt changes. \textit{Job searching}'s popularity decreased sharply, while \textit{Interviews} and \textit{Leaving a job} increased their popularity. After April 2021 the share of the flairs did not reverse to the pre-pandemic trends. Instead \textit{Job searching}' share remained lower than in pre-pandemic levels and the share of \textit{Job offers} showed a steady increase. 

The average share of posts flaired under \textit{Leaving a job} after April 2021 is roughly equal to the pre-pandemic average share. This finding might be surprising given the large number of quits that happened during the Great Resignation. One must bear in mind, however, that some posts flaired under \textit{Leaving a job} may correspond to fires, while posts about quitting may have a different flair. For example, people may mention they are quitting their job to take a different job and flair the post with \textit{Job offers}. However, as we showed in the main text, when we distinguish posts as  quit-related and fired-related, we find that the number of quit-related posts increased during the Great Resignation, while the fired-related posts spiked in the first months of the pandemic.

\FloatBarrier

\section{Sentiment analysis}\label{sec:SI_sentiment}
Here we present additional results and robustness checks for the sentiment analysis.

\subsection{Sentiment analysis in `r/jobs' with NRC emotion lexicon}

\begin{figure}[h]
    \centering
\includegraphics[width = 0.95\textwidth]{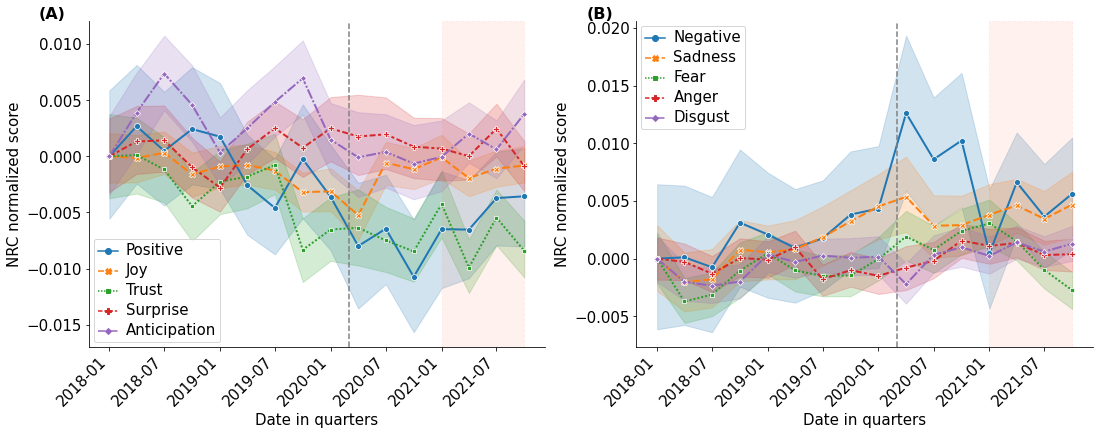}
    \caption{\textbf{Sentiment analysis} This plot shows the dynamics of the sentiment score of 'r/jobs' posts. (A) Sentiments that are considered positive or neutral (B) Sentiments that are considered negative.}
    \label{fig:sentiment_all_quarters}
\end{figure}

In Figure~ \ref{fig:sentiment_all_quarters} we present the overall NRC \cite{mohammad2013crowdsourcing} sentiment score, for both polarities and feelings, across quit and non-quit related posts. Panel A shows positive polarity and positive or neutral feelings. Panel B shows negative polarity and negative feelings. Before the pandemic the sentiment expressed on `r/jobs' was deteriorating. Negative feelings such as fear, sadness and disgust were trending upwards, while joy and trust were downward trending. Negative polarity reaches record high levels in the early months of the pandemic, while positive polarity reaches record low levels. However, these changes are short-lived. We alsl notice that fear and anger scores have been declining since the beginning of 2021, coinciding with the Great Resignation period, while joy and trust stabilized since mid 2020. Overall, we see a reversal of trends since the start of the Great Resignation - upward trending overall positive sentiment, and downward trending overall negative sentiment.

\begin{figure}[h]
    \centering
    \includegraphics[width = 0.95\textwidth]{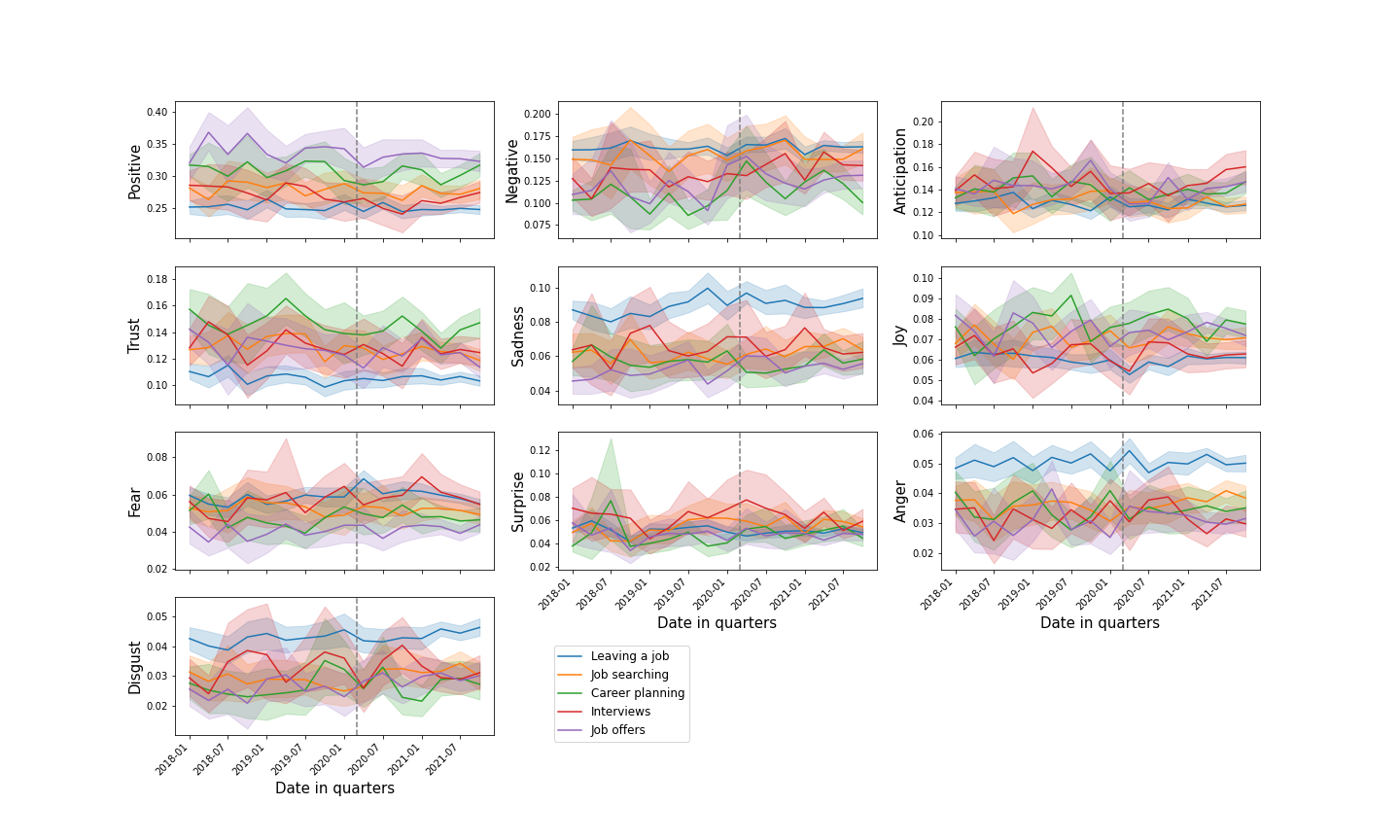}
    \caption{\textbf{Sentiment analysis} This plot shows the average sentiment score across quarters for posts of different flairs. The grey line marks first of March 2020. Only posts from the quit and control group are included}
    \label{fig:sentiment_flair_quarters}
\end{figure}

\paragraph{Sentiment across flairs}
To understand how the flair of a post relates to sentiment we look into the average sentiment for each of the top five flairs. We do this using all filtered posts and the NRC dictionary \cite{mohammad2013crowdsourcing}. We find that across the top five flairs Positive sentiment scores at the highest (see Figure~\ref{fig:sentiment_flair_quarters}). This is somewhat surprising given that some flairs, such as \textit{Leaving a job}, may be intuitively more linked with negative sentiment. Nonetheless, we find that the relative scores between posts of different flairs are intuitive. For example, posts flaired \textit{Leaving a job} 
score high in Sadness, Anger, and Disgust. In contrast, \textit{Job offers} posts 
score higher in Positive sentiment and lower in Fear. 
For the sentiment analysis in the main text we use a difference in differences approach that focuses on changes in sentiment scores before and after the pandemic. Since the NRC dictionary \cite{mohammad2013crowdsourcing} captures relative differences between the flairs that are intuitive, we consider that the NRC dictionary \cite{mohammad2013crowdsourcing} is an appropriate method to perform sentiment analysis. Nonetheless, for robustness, we look further into the NRC dictionary \cite{mohammad2013crowdsourcing} below and use other sentiment analysis methods later in this appendix.

\paragraph{Understanding NRC results}
We looked further into what drove the results of NRC scores by looking at the posts with the highest negative scores. As the NRC gives a distribution of sentiments in each document, a negative score of one means only the negative sentiment was found. We took a random sample of 30 posts and looked closer at the words which triggered the negative score. Some of the words such as `wrong' or `tired' were coherent with our own expectations but some others, such as `small', `minimum', or `foreign' were more questionable. Additionally, the word `quit', particularly important in the present study, was classified as negative, which limits the possibility of quitting being considered as a positive and emancipating event (i.e., as described in \cite{Goldberg_2021}).
We also looked into the words that were classified as positive. We find that both `working' and `job' are considered positive. These words explain why the predominant sentiment is positive. We do not think this results hinders the results of our analysis, since we focus on differences. Nonetheless, we run robustness checks with alternative sentiment methods.

\subsection{Difference-in-differences analysis and alternative methods for sentiment analysis}
\begin{figure}[h]
    \centering
    \includegraphics[width = 0.75\textwidth]{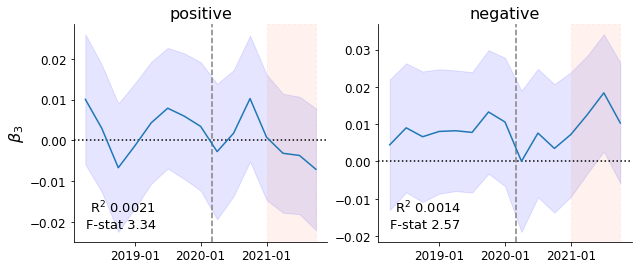}
    \caption{Difference-in-Differences analysis between quit and non-quit posts. Results for positive and negative polarity using NRC. }
    \label{fig:app_sent_posneg}
\end{figure}

Figure~\ref{fig:app_sent_posneg} shows the results for the difference-in-differences analysis for NRC polarity scores. As discussed in the main text, the positive and negative polarity are noisy and do not show particularly interesting or significant trends. To further validate our results we use two different sentiment analysis methods.  

\begin{figure}[h]
    \centering
    \includegraphics[width = 0.65\textwidth]{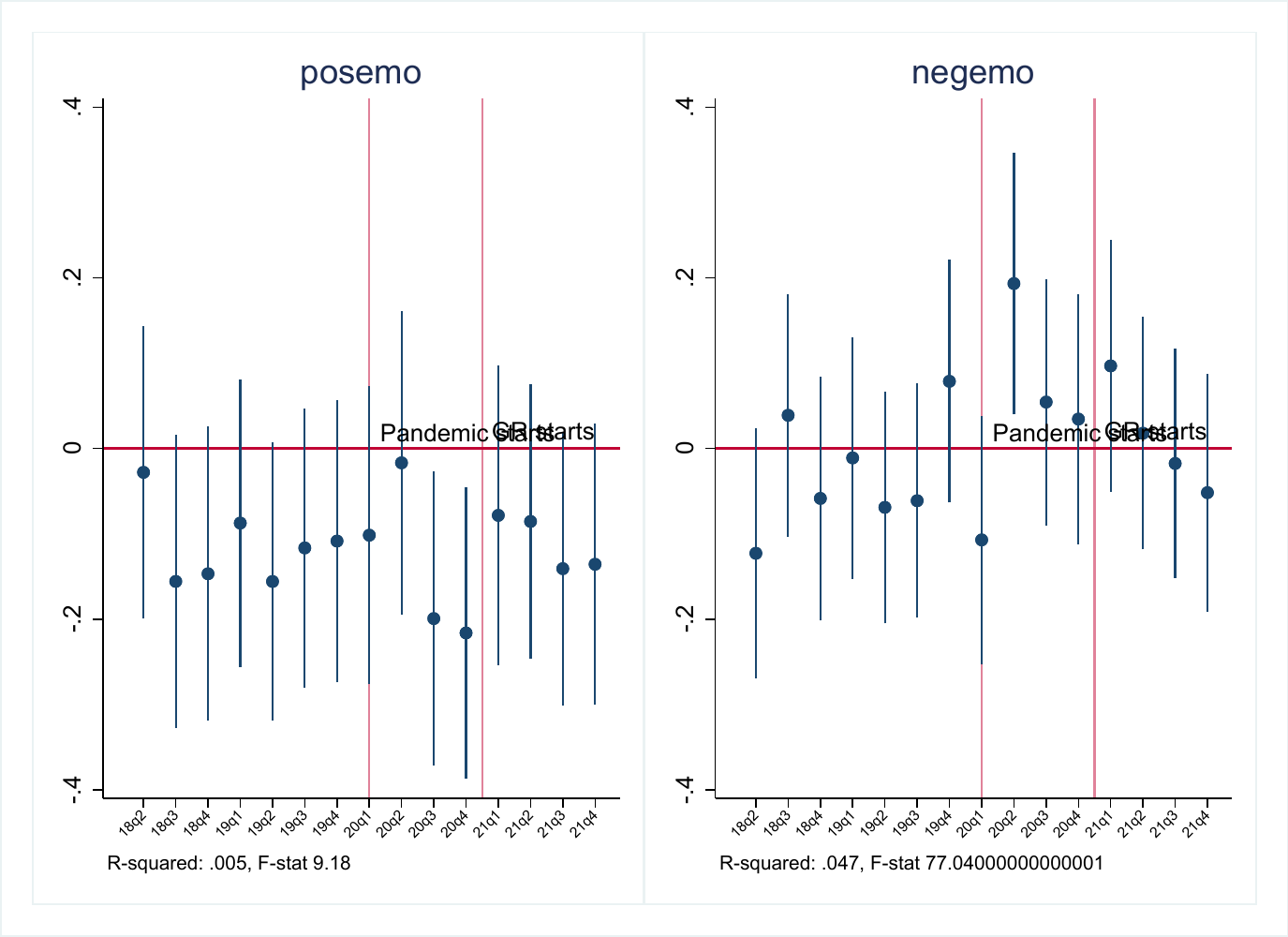}
\includegraphics[width = 0.75\textwidth]{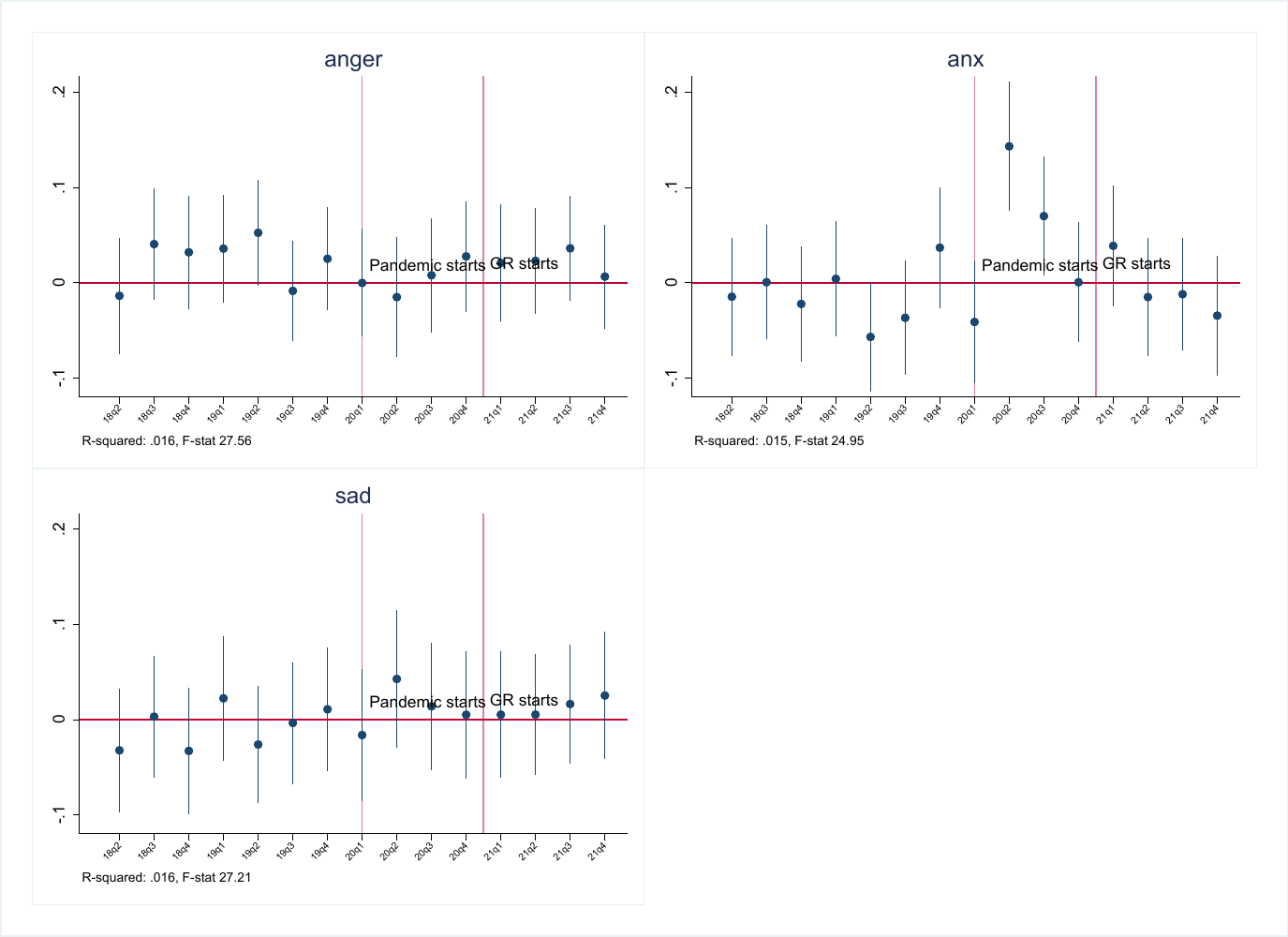}
    \caption{\textbf{LIWC sentiment analysis} Difference in differences analysis for positive and negative emotion (top panels) and anxiety, anger and sadness (bottom panels). The first red line from left to right marks the start of the pandemic, the second red line marks the Great Resignation period starting}
    \label{fig:sentiment_liwc}
\end{figure}

\paragraph{Different methods for sentiment analysis}
 We use a different dictionary based approach, LIWC \citep{pennebaker2001linguistic}. We measure the sentiment of the affective processes (positive emotion, negative emotion, anxiety, anger, and sadness) for the quit and control group and perform a difference in differences analysis as in the main text (see Figure Figure~\ref{fig:sentiment_liwc}). Our results show that negative emotion and anxiety increased significantly among quit-related posts relative to non-quit posts during the first (and for anxiety also second) quarter of the pandemic. The rest of the sentiments did not show any significant change. 
 
\begin{figure}[h]
    \centering
    \includegraphics[width = 0.65\textwidth]{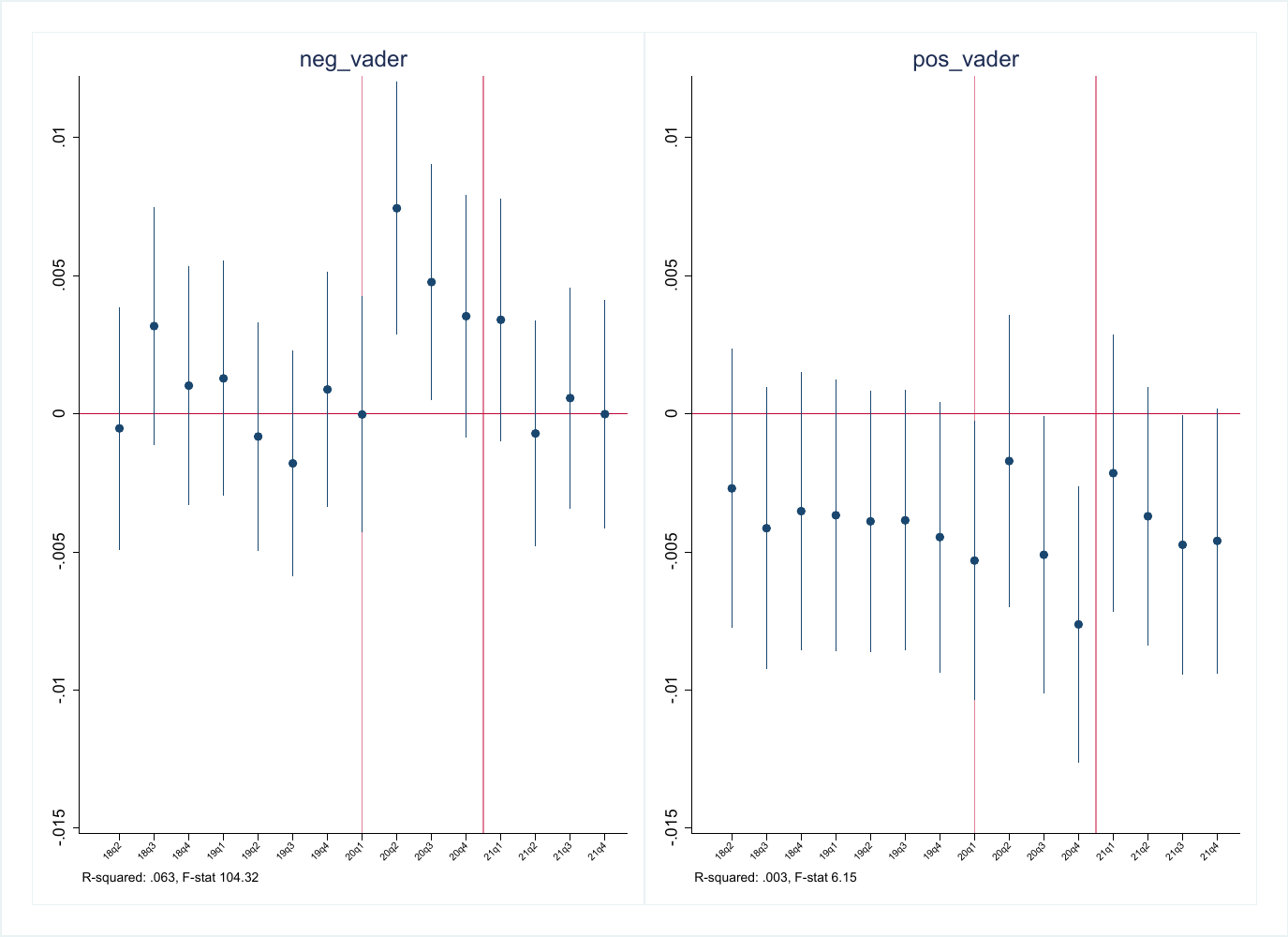}
    \caption{\textbf{Vader sentiment analysis} Difference in differences analysis for positive and negative sentiment. The first red line from left to right marks the start of the pandemic, the second red line marks the Great Resignation period starting}
    \label{fig:sentiment_vader}
\end{figure}

As a second robustness check on the sentiment analysis, we use VADER~\citep{hutto2014vader}, which is particularly developed for social media analysis. When using VADER we do not go through the pre-processing steps of lemmatization and negation handling, since VADER does not require these steps. The results are shown in Figure~\ref{fig:sentiment_vader}. We see a significant increase in negative sentiment across quit-related posts relative to non-quit related posts in the first two quarters of the pandemic. Once again, this effect is not long-lasting, as negative sentiment returns back to pre-pandemic levels after the second quarter of the pandemic. 

\FloatBarrier
\section{Topic modelling} \label{sec:SI_topicmodelling}
In this section we present the results for all topics from our topic modelling analysis. Furthermore, we present additional analysis for some topics of interest.

\begin{figure}[h]
    \centering
\includegraphics[width = 0.9\textwidth]{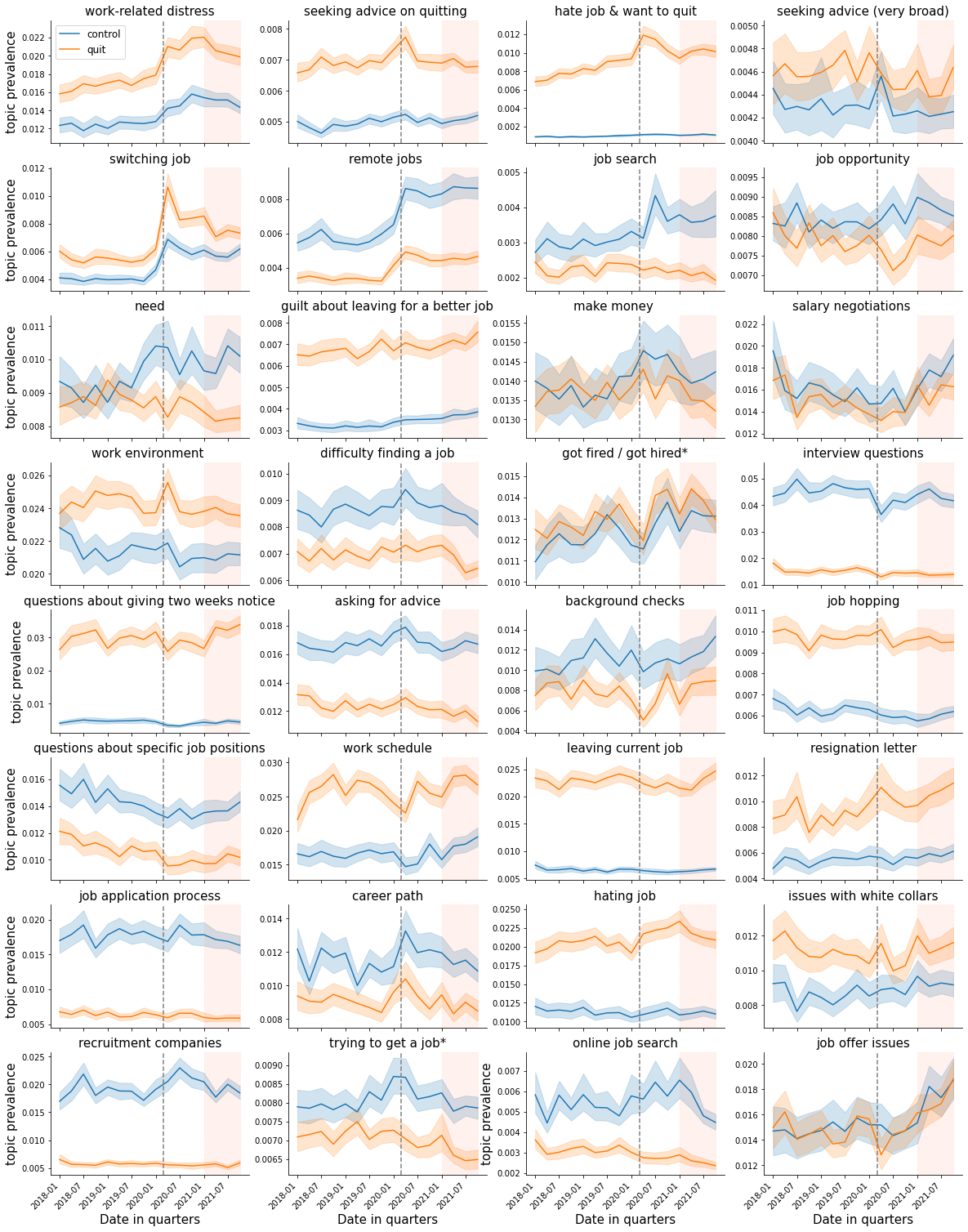}
    \caption{\textbf{Topics prevalence} This plot shows the dynamics of the prevalence of different topics}
    \label{fig:topic_prevalence_apx_1}
\end{figure}

\begin{figure}[h]
    \centering
\includegraphics[width = 0.9\textwidth]{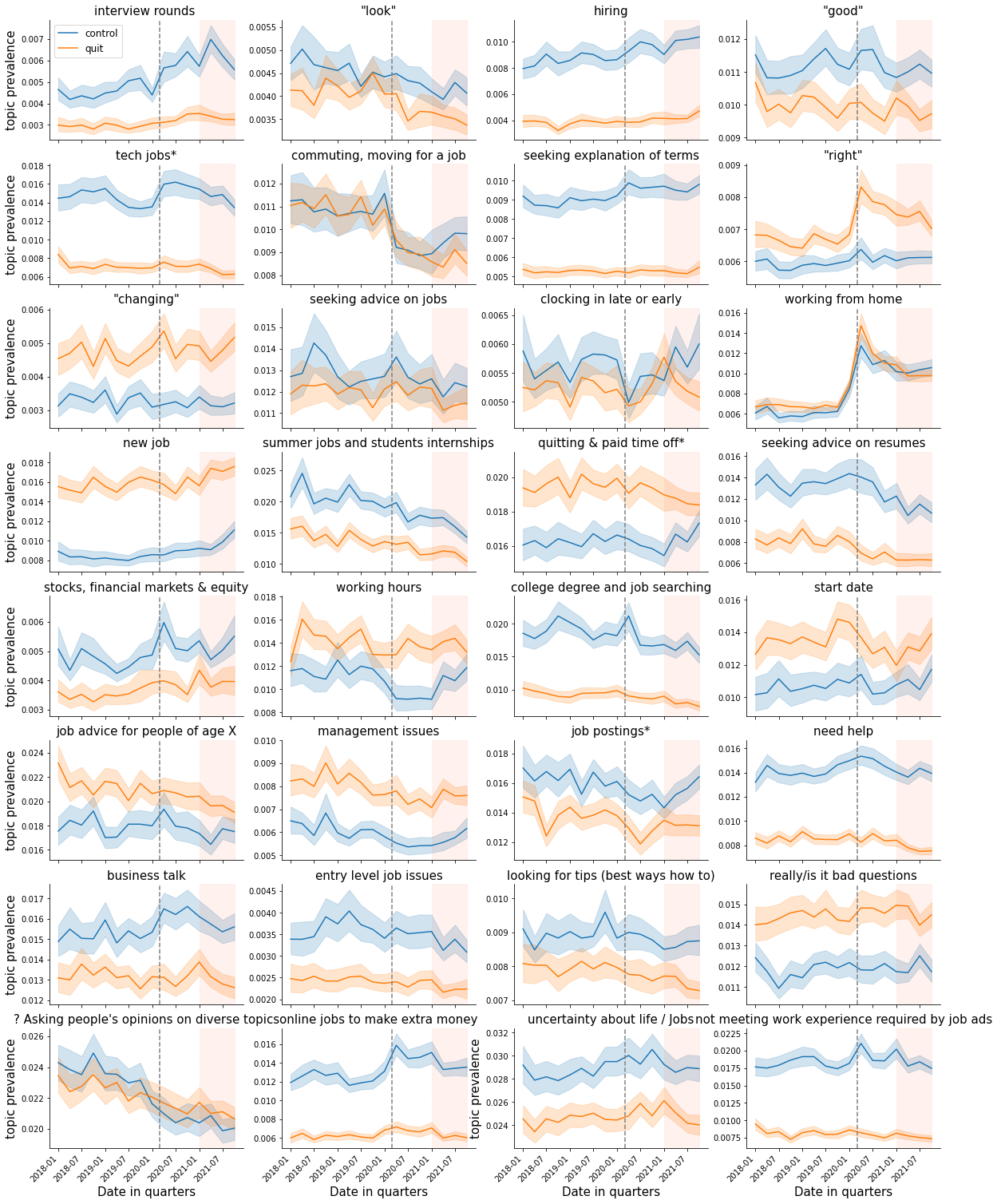}
    \caption{\textbf{Topics prevalence} This plot shows the dynamics of the prevalence of different topics}
    \label{fig:topic_prevalence_apx_2}
\end{figure}

In Table~\ref{tab:si_topc} we show the names of all topics and whether they are clear topics (CT), multi-topics (MT), or boiler-plate topics (BT). Figures \ref{fig:topic_prevalence_apx_1} -- \ref{fig:topic_prevalence_apx_2} show the dynamics of the prevalence of all 90 topics across quit and non-quit related posts. Table \ref{tab:si_topc_did} shows the results of the difference in differences analysis for 78 interpretable topics.

\begin{figure}[h]
    \centering
\includegraphics[width = 0.9\textwidth]{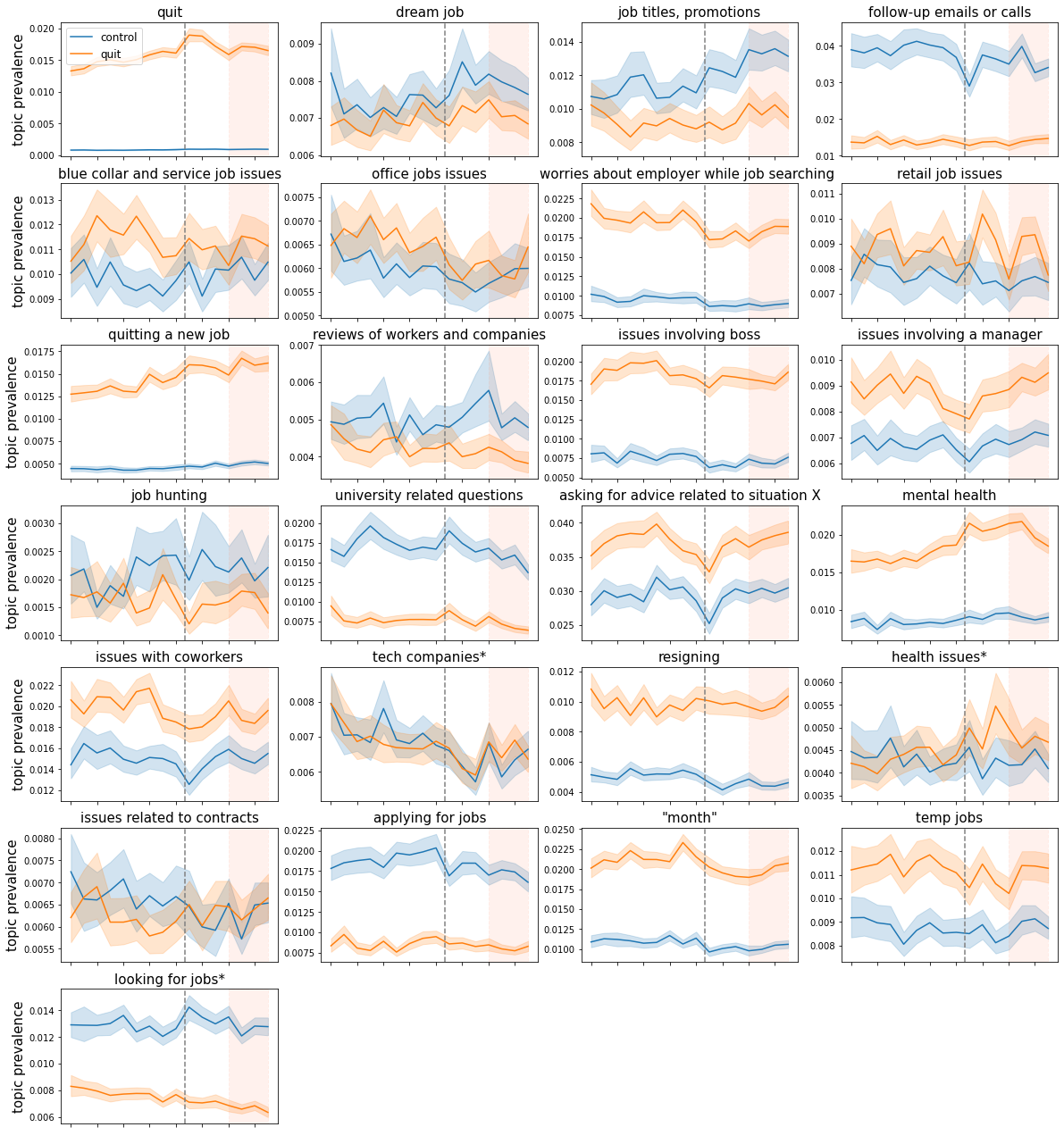}
    \caption{\textbf{Topics prevalence} This plot shows the dynamics of the prevalence of different topics}
    \label{fig:topic_prevalence_apx_3}
\end{figure}

\begin{figure}[h!]
   \centering
    \includegraphics[width = 0.75\textwidth]{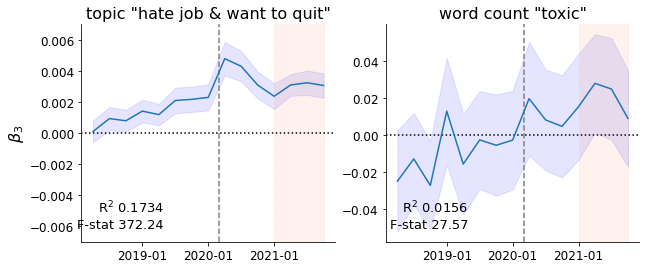}
    \caption{Difference-in-Differences. Relative changes in prevalence in selected topics among quit-related posts}
    \label{fig:app_topic_toxic}
\end{figure}

To understand the drivers of the dynamics of some topics mentioned in the main text, we did an additional analysis using word counts. The difference-in-differences analysis for the topic \textit{hate job \& want to quit} showed an increasing pre-trend before the pandemic (see left panel of Figure~\ref{fig:app_topic_toxic}). To better understand what drives this pre-trend we also performed a difference-in-differences analysis on the word count of the word `toxic'. As the right panel of Figure~\ref{fig:app_topic_toxic} shows, people started using the word `toxic' more in quit than in non-quit related posts increasingly before the pandemic. Therefore, as we mention in the main text, the rise of concerns about toxic work and quitting cannot be fully attributed to the pandemic. 

\begin{figure}[h!]
   \centering
    \includegraphics[width = 0.75\textwidth]{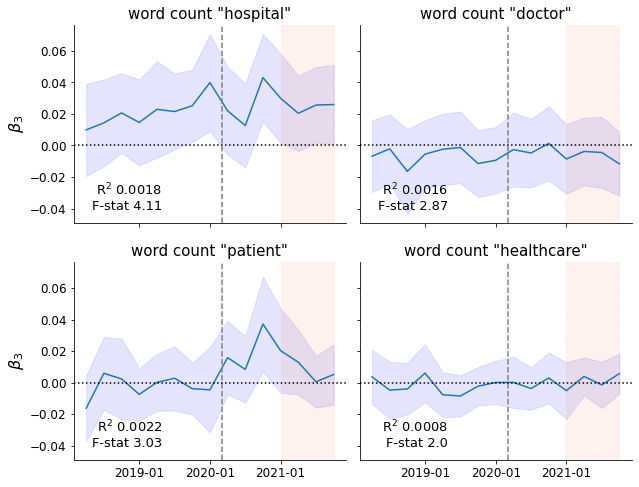}
    \caption{Difference-in-Differences. Relative changes in prevalence in selected topics among quit-related posts}
    \label{fig:app_topic_healthcare}
\end{figure}

In the main text we also mention an increase in the prevalence of the multi-topic \textit{health issues / healthcare job / scheduling} in quit related posts with respect to non-quit posts. To try to understand which of the subtopics was driving the increase we also did a difference-in-differences analysis on different common words among the topic. We were not able to find a particular group of words to which we could clearly attribute the increase of the topic prevalence (see Figure~\ref{fig:app_topic_healthcare} for examples of words we looked into). As we conclude in the main text, we are not able to identify precisely what drives the increase of this topic and hence our results for this topic should be nuanced.

\paragraph{Robustness check}
To verify the robustness of our main result, i.e., the shifts on the mental health discourse, we repeat the topic modelling analysis varying the number of topics $K$. We use one lower bound and one upper bound $K=70$ and $K=110$ respectively. While both solutions deliver a different set of topics (less and more exclusive respectively), we observe that our main results are mirrored in them. In the solution at $K=70$, the term \textit{mental health} is included in Topic 27, where it is interwined with the terms idiosyncratic to the topic \textit{hating job} in our solution at $K=90$ (FREX: \textit{hate, anxiety, miserable, stress, suck, absolutely, depression, tired, hate job,} and \textit{mental health}. This topic shows both an increase across time in a model estimating eq. (\ref{eq:DiD_topic_basic}) ($\beta_1$ = .002, $p <$ .000), as well as a significant interaction effect in a model estimating eq. (\ref{eq:DiD_simple}) ($\beta_3$ = .003, $p <$ .000). In the solution with $K=110$, \textit{mental health} is included in Topic 80 (the same position as in our solution at $K=90$) with almost fully overlapping Top 10 FREX keywords. The results are similar in both the simple ($\beta_1$ = .002, $p <$ .000) and moderated ($\beta_3$ = .003, $p <$ .000) models. The topic \textit{work-related distress} is found in the same position (Topic 1) at $K=70$ and $K=90$ and as Topic 108 at $K=110$, and is represented by mostly identical Top 10 FREX keywords in both models. Computing overall time changes with eq. (\ref{eq:DiD_topic_basic}) and changes idiosyncratic to quit posts with eq. (\ref{eq:DiD_simple}) shows positive and statistically significant increases in both models. Other relevant topics (e.g., \textit{hate job \& want to quit}) are also found in the models at $K=70$ and $K=110$ and show qualitatively identical results across time.

\begin{longtable}{c|c|p{0.3\linewidth}|p{0.5\linewidth}}
\caption{\textbf{Labels for the 90 topics and their keywords}. CL: clear topics; MT: multi-topics; BT: boiler-plate topics}\\
\hline
\multicolumn{1}{c|}{\textbf{Topic}} &
\multicolumn{1}{c|}{\textbf{Type}} &
\multicolumn{1}{c|}{\textbf{Label}} & \multicolumn{1}{c}{\textbf{Keywords}} \\ \hline 
\endfirsthead

\multicolumn{4}{c}%
{{\bfseries \tablename\ \thetable{} -- continued from previous page}} \\
\hline \multicolumn{1}{c|}{\textbf{Topic}} &
\multicolumn{1}{c|}{\textbf{Type}} &
\multicolumn{1}{c|}{\textbf{Label}} & \multicolumn{1}{c}{\textbf{Keywords}} \\
\endhead

\multicolumn{4}{r}{{Continued on next page}} \\ 
\endfoot
\endlastfoot

        1 & CL & Work-related distress & feel\_like, feel, dont\_feel, just\_feel, feeling, lost, job\_feel, like\_just, completely, feeling\_like \\ \hline
        2 & MT & seeking advice on quitting  & quitting\_job, unemployment, okay, short, half, text, paycheck, regret, possibly, story \\ \hline
        3 & CL & hate job \& want to quit & quit\_job, toxic, want\_quit, just\_quit, properly, toxic\_work, workplace, quit\_current, quit\_dont, shitty \\ \hline
        4 & BT & ``worth" / ``soon" & worth, possible, considered, changing\_job, soon, soon\_possible, dilemma, promised, job\_soon, considering \\ \hline
        5 & CL & switching job & pandemic, company\_year, current\_company, laid, working\_company, employed, large, switching, large\_company, switch\_job \\ \hline
        6 & CL & remote jobs  & remote, support, accounting, remotely, fully, country, remote\_job, relocate, video, consultant \\ \hline
        7 & CL & job search & job\_searching, searching, searching\_job, nyc, unemployed, shot, actively, rough, turned, turning \\ \hline
        8 & CL & job opportunity & opportunity, great, growth, job\_opportunity, grow, amazing, great\_job, board, new\_opportunity, awesome \\ \hline
        9 & BT & ``need" & need\_job, really\_need, just\_need, job\_need, need\_advice, need, dont\_need, need\_money, break, need\_work \\ \hline
        10 & CL & guilt about leaving for a better job & leaving\_job, better\_job, feel\_guilty, better, guilty, job\_better, better\_pay, burn, bridge, guilt \\ \hline
        11 & CL & make money & make, make\_sure, want\_make, make\_money, money, sense, make\_sense, sure, making, job\_make \\ \hline
        12 & CL & salary negotiations & salary, bonus, raise, increase, higher, negotiate, base, range, lower, compensation \\ \hline
        13 & CL & work environment  & want\_work, work\_environment, job\_work, looking\_work, like\_work, place\_work, work\_just, environment, work, currently\_work \\ \hline
        14 & CL & difficulty finding a job & finding, finding\_job, hard\_time, grad, hard, recent, difficult, trouble, having, harder \\ \hline
        15 & MT & got fired / got hired & just\_got, got\_job, got, job\_got, got\_fired, recently\_got, got\_hired, fired, finally\_got, finally \\ \hline
        16 & CL & interview questions & job\_interview, answer, interviewer, question, interview, answer\_question, phone\_interview, interview\_tomorrow, second\_interview, interview\_job \\ \hline
        17 & CL & questions about giving two weeks notice & week\_notice, notice, vacation, notice\_period, week, giving, giving\_notice, job\_week, period, giving\_week \\ \hline
        18 & CL & asking for advice & advice, advice\_appreciated, thanks\_advance, thanks, appreciate, appreciated, greatly\_appreciated, advance, situation, job\_advice \\ \hline
        19 & CL & background checks & check, background\_check, background, record, employment, drug\_test, paperwork, completed, credit, worried \\ \hline
        20 & CL & job hopping & stay, long, story\_short, long\_story\_short, want\_stay, long\_term, staying, longer, term, long\_time \\ \hline
        21 & CL & questions about specific job positions & position, current\_position, offered\_position, new\_position, job\_position, position\_company, offered, position\_year, different, open \\ \hline
        22 & CL & work schedule & day\_week, day, work\_day, weekend, shift, day\_day, night, day\_work, job\_day, saturday \\ \hline
        23 & CL & leaving current job & leave\_job, current\_job, want\_leave, leave, current, leave\_current, just\_leave, job\_current, dont\_want, want \\ \hline
        24 & CL & (resignation) letter & letter, resignation, signed, sign, resignation\_letter, offer\_letter, written, formal, write, recommendation \\ \hline
        25 & CL & job application process & application, job\_application, address, applied\_job, applicant, paper, form, number, applied, status \\ \hline
        26 & CL & career path & change, career, path, choice, career\_path, industry, change\_job, choose, goal, direction \\ \hline
        27 & CL & hating job & hate, absolutely, suck, miserable, dont\_like, hate\_job, everyday, tired, anymore, shit \\ \hline
        28 & CL & issues with white collars & team, project, workload, lead, member, leader, department, deadline, team\_member, leadership \\ \hline
        29 & CL & recruitment companies & linkedin, recruiter, website, bank, profile, post, site, information, recruitment, recruiting \\ \hline
        30 & MT & trying to get a job / looking for a job for relatives & trying, figure, understand, dad, trying\_job, foot, mom, door, trying\_figure, dont\_understand \\ \hline
        31 & CL & online job search & job\_search, search, wife, realize, google, networking, network, resource, unique, connection \\ \hline
        32 & CL & job offer issues & job\_offer, offer, offer\_company, accepted, accepted\_job, offer\_job, accepting, got\_offer, received\_offer, accept \\ \hline
        33 & CL & interview rounds  & land, round\_interview, round, mcdonalds, final, land\_job, managed, hard\_work, work\_hard, landed \\ \hline
        34 & BT & ``look" & look\_like, look\_job, look\_bad, look, job\_look, make\_look, like\_job, look\_good, job\_job, eye \\ \hline
        35 & CL & hiring  & hiring, hiring\_manager, process, hiring\_process, interview\_process, candidate, stage, onboarding, red\_flag, screening \\ \hline
        36 & BT & ``good" & good\_job, good, job\_good, really\_good, fit, good\_fit, pretty\_good, pretty, good\_idea, think\_good \\ \hline
        37 & MT & tech jobs / tests in hiring process & test, software, engineer, engineering, design, technical, learn, hotel, learning, technician \\ \hline
        38 & CL & commuting, moving for job & car, location, city, drive, driving, house, town, commute, mile, live \\ \hline
        39 & CL & seeking explanation of terms & doe, mean, sound, sound\_like, anybody, director, doe\_mean, usually, proceed, weird \\ \hline
        40 & BT & ``right" & right, job\_right, mess, hell, away, right\_away, dumb, right\_thing, sorry, idiot \\ \hline
        41 & BT & ``changing" & changing, exact, jump, keep, switched, ground, catch, wonder, thing\_just, thing \\ \hline
        42 & BT & ``high" & paying, high, job\_pay, bill, high\_school, low, paying\_job, decent, debt, pay \\ \hline
        43 & CL & clocking in late or early & late, early, met, meet, minute, fault, suppose, vent, missing, strange \\ \hline
        44 & CL & working from home & home, work\_home, family, working\_home, kid, disability, covid, child, baby, family\_member \\ \hline
        45 & CL & new job & new\_job, new, job\_new, started\_new, new\_company, starting\_new, looking\_new, starting, got\_new, start\_new \\ \hline
        46 & CL & summer jobs and student internships & summer, internship, school, college, semester, class, college\_student, intern, job\_college, student \\ \hline
        47 & MT & quitting \& paid time off / considering part time jobs & time\_job, time, working\_time, paid\_time, work\_time, time\_time, waste, time\_work, time\_just, spend \\ \hline
        48 & CL & seeking advice on resumes & gap, resume, list, cover\_letter, include, history, job\_resume, work\_history, explain, lie \\ \hline
        49 & CL & stocks, financial markets \& equity & market, job\_market, value, welcome, min, stock, share, heavily, option, labor \\ \hline
        50 & CL & working hours & hour\_week, parttime, fulltime, hour, working\_hour, parttime\_job, work\_hour, second\_job, fulltime\_job, hour\_work \\ \hline
        51 & CL & college degree and job searching & field, bachelor, degree, bachelor\_degree, job\_field, graduated, master\_degree, master, certification, education \\ \hline
        52 & CL & start date & reference, start, start\_date, june, july, date, april, job\_start, january, august \\ \hline
        53 & CL & job advice for people of age X & old, year\_old, job\_year, old\_job, past\_year, year, year\_ago, worked\_year, year\_year, year\_job \\ \hline
        54 & CL & management issues & employee, staff, nonprofit, force, organization, husband, rule, corporate, new\_employee, management \\ \hline
        55 & MT & job postings / comparing two options / companies & job\_posting, job\_company, work\_company, company\_just, company\_company, posting, big\_company, company, company\_want, company\_work \\ \hline
        56 & CL & need help & help, need\_help, thank, guy, reddit, hello, construction, hey, job\_help, thank\_advance \\ \hline
        57 & BT & business talk & business, people, lot\_people, small\_business, people\_work, youre, like\_people, usa, brand, arent \\ \hline
        58 & CL & entry level job issues & entry\_level, entry\_level\_job, stuck, dead, entry\_level\_position, mid, progress, security, job\_doing, level \\ \hline
        59 & CL & looking for tips (best ways how to) & whats, best, way, best\_way, whats\_best, approach, youve, suit, rep, worst \\ \hline
        60 & BT & ``really" / ``is it bad" questions & really\_like, feel\_bad, bad, really\_want, really, job\_really, really\_bad, dont\_really, bad\_idea, idea \\ \hline
        61 & BT & ``think" & quite, dont\_think, think, sort, real, bit, essentially, somewhat, kind, probably \\ \hline
        62 & CL & online jobs to make extra money & free, marketing, earn, amazon, teacher, teaching, online, teach, writing, social\_media \\ \hline
        63 & BT & uncertainty about life / jobs & dont\_know, know, want\_know, know\_job, let\_know, let, want\_job, job\_know, job\_dont, know\_just \\ \hline
        64 & CL & not meeting work experience required by job ads & work\_experience, year\_experience, experience, skill, relevant, qualification, volunteer, job\_experience, lack, require \\ \hline
        65 & CL & quit & quit, job\_quit, today, job\_today, spot, quit\_week, quit\_just, scared, understaffed, wanting \\ \hline
        66 & CL & dream job & dream, dream\_job, love, wanted, rejected, loved, just\_wanted, love\_job, wanted\_work, really\_wanted \\ \hline
        67 & CL & job titles, promotions & title, role, description, job\_title, responsibility, job\_description, promotion, title\_say, new\_role, current\_role \\ \hline
        68 & CL & follow-up emails or calls & follow, sent, email, emailed, monday, heard, friday, saying, havent\_heard, phone \\ \hline
        69 & CL & blue collar and service job issues & warehouse, owner, order, food, worker, service, cleaning, kitchen, server, table \\ \hline
        70 & CL & office jobs issues & office, office\_job, admin, desk, sit, assistant, receptionist, sitting, space, duty \\ \hline
        71 & CL & worries about employer while job searching & leaving, employer, previous, previous\_job, reason, left\_job, potential\_employer, left, current\_employer, previous\_employer \\ \hline
        72 & CL & retail job issues & retail, store, retail\_job, fast\_food, customer, restaurant, cashier, seasonal, grocery\_store, customer\_service \\ \hline
        73 & CL & quitting a new job & quitting, just\_started, started, started\_working, thinking, thinking\_quitting, job\_just, recently\_started, started\_job, job\_started \\ \hline
        74 & CL & reviews of workers and companies & review, performance, negative, positive, exam, feedback, poor, probation, improve, glassdoor \\ \hline
        75 & CL & issues involving boss & bos, shes, boss, tell, tell\_bos, told\_bos, upset, current\_bos, coworkers, bos\_said \\ \hline
        76 & CL & issues involving a manager & rant, manager, assistant\_manager, new\_manager, promoted, manager\_told, manager\_said, general, told\_manager, manager\_position \\ \hline
        77 & CL & job hunting & hunting, job\_hunting, hunt, job\_hunt, rejection, wasnt, march, start\_job, luckily, decided \\ \hline
        78 & CL & university related questions & program, university, study, research, finance, international, studying, uni, phd, math \\ \hline
        79 & CL & asking for advice related to situation X & didnt, said, wasnt, knew, told, kept, called, asked, didnt\_want, came \\ \hline
        80 & CL & mental health & anxiety, mental\_health, lined, stress, job\_lined, depression, mental, worse, physical, mentally \\ \hline
        81 & CL & issues with coworkers & supervisor, coworker, meeting, woman, mistake, group, colleague, task, uncomfortable, talk \\ \hline
        82 & MT & tech companies / small vs large companies & switch, tech, firm, small\_company, consulting, pro, con, small, tech\_company, smaller \\ \hline
        83 & CL & resigning & resign, concern, responsible, case, claim, terminated, forced, agreement, resigned, termination \\ \hline
        84 & MT & health issues / jobs in healthcare / scheduling issues & hospital, medical, patient, doctor, appointment, scheduling, request, requested, facility, healthcare \\ \hline
        85 & CL & issues related to contracts & ceo, agency, client, contractor, insurance, contract, experienced, package, agent, payroll \\ \hline
        86 & CL & applying for jobs & applying, applying\_job, apply, getting\_job, chance, apply\_job, job\_applying, gotten, getting, chance\_getting \\ \hline
        87 & BT & ``month" & month\_ago, job\_month, couple\_month, month, quit\_month, past\_month, ago, month\_just, month\_later, month\_month \\ \hline
        88 & CL & temp jobs & hire, training, temp, hired, train, week\_ago, temporary, couple\_week, permanent, new\_hire \\ \hline
        89 & MT & looking for jobs / sales jobs related questions & looking\_job, sale, looking, suggestion, start\_looking, job\_looking, startup, sell, just\_looking, product \\ \hline
        90 & MT & asking for advice (broad)  & regardless, going, ill, turn, basically, come, thought, wont, there, actually \\ \hline
\label{tab:si_topc}
\end{longtable}

\begin{longtable}{c|c|p{0.25\linewidth}|p{0.15\linewidth}|p{0.15\linewidth}|p{0.15\linewidth}}
\caption{\textbf{Difference in differences analysis for the 78 interpretable topics}. CL: clear topics; MT: multi-topics; and BT: boiler plate topics}\\
\hline
\multicolumn{1}{c|}{\textbf{Topic}} &
\multicolumn{1}{c|}{\textbf{Type}} &
\multicolumn{1}{c|}{\textbf{Label}} & \multicolumn{1}{c|}{\textbf{Parallel trends?}}& \multicolumn{1}{c|}{\textbf{R-squared}} & \multicolumn{1}{c}{\textbf{F-statistic}} \\ \hline 
\endfirsthead

\multicolumn{6}{c}%
{{\bfseries \tablename\ \thetable{} -- continued from previous page}} \\
\hline \multicolumn{1}{c|}{\textbf{Topic}} &
\multicolumn{1}{c|}{\textbf{Type}} &
\multicolumn{1}{c|}{\textbf{Label}} & \multicolumn{1}{c|}{\textbf{Parallel trends?}}& \multicolumn{1}{c|}{\textbf{R-squared}} & \multicolumn{1}{c}{\textbf{F-statistic}} \\ \hline 
\endhead

\multicolumn{6}{r}{{Continued on next page}} \\ 
\endfoot
\endlastfoot

    1     & CL    & work-related distress & yes   & 0.027 & 46.3 \\
    2     & MT    & seeking advice on quitting  & yes   & 0.051 & 92.19 \\
    3     & CL    & hate job \& want to quit & no    & 0.173 & 372.24 \\
    5     & CL    & switching job & nearly yes & 0.028 & 46.43 \\
    6     & CL    & remote jobs  & yes   & 0.034 & 57.54 \\
    7     & CL    & job searching & nearly yes & 0.007 & 13.09 \\
    8     & CL    & job opportunity & no    & 0.002 & 4.24 \\
    10    & CL    & guilt about leaving for a better job & nearly yes & 0.058 & 112.17 \\
    11    & CL    & make money & no    & 0     & 1.55 \\
    12    & CL    & salary negotiations & no    & 0.002 & 4.12 \\
    13    & CL    & work environment  & no    & 0.007 & 13.49 \\
    14    & CL    & difficulty finding a job & yes   & 0.007 & 13.6 \\
    15    & MT    & got fired/got hired & no    & 0.002 & 4.25 \\
    16    & CL    & interview questions & no    & 0.085 & 156.65 \\
    17    & CL    & questions about giving two weeks notice & no    & 0.094 & 182.81 \\
    18    & CL    & asking for advice & yes   & 0.029 & 53.24 \\
    19    & CL    & background checks & no    & 0.003 & 8.42 \\
    20    & CL    & job hopping & yes   & 0.045 & 81.15 \\
    21    & CL    & questions about specific job positions & yes   & 0.014 & 25.38 \\
    22    & CL    & work schedule & no    & 0.025 & 44.3 \\
    23    & CL    & leaving current job & yes   & 0.122 & 239.57 \\
    24    & CL    & resignation letter & yes   & 0.012 & 20.79 \\
    25    & CL    & job application process & nearly yes & 0.056 & 102.62 \\
    26    & CL    & changing career path & no    & 0.007 & 12.27 \\
    27    & CL    & hating job & no    & 0.051 & 93.33 \\
    28    & CL    & issues with white collars & yes   & 0.005 & 9.24 \\
    29    & CL    & recruitment companies & no    & 0.085 & 159.28 \\
    30    & MT    & trying to get a job / looking for a job for relatives & yes   & 0.007 & 13.35 \\
    31    & CL    & online job search & yes   & 0.014 & 26.71 \\
    32    & CL    & job offer issues & yes   & 0.002 & 3.77 \\
    33    & CL    & interview rounds  & yes   & 0.019 & 28.78 \\
    35    & CL    & hiring  & nearly yes & 0.038 & 68.8 \\
    37    & MT    & tech jobs* & no    & 0.042 & 75.09 \\
    38    & CL    & commuting, moving for job & yes   & 0.003 & 7.43 \\
    39    & CL    & seeking explanation of terms & yes   & 0.053 & 96.86 \\
    43    & CL    & clocking in late/early & yes   & 0.001 & 3.22 \\
    44    & CL    & working from home & yes   & 0.024 & 41.9 \\
    45    & CL    & new job & no    & 0.04  & 74.32 \\
    46    & CL    & summer jobs / internships for college students  & nearly yes & 0.017 & 28.29 \\
    47    & MT    & quitting \& paid time off | considering part time jobs & no    & 0.008 & 14.96 \\
    48    & CL    & seeking advice on resumes & yes   & 0.021 & 37.92 \\
    49    & CL    & stocks, financial markets \& equity & no    & 0.006 & 11.4 \\
    50    & CL    & working hours & no    & 0.005 & 10.85 \\
    51    & CL    & college degree and job searching & no    & 0.033 & 56.7 \\
    52    & CL    & start date & yes   & 0.005 & 9.3 \\
    53    & CL    & job advice for people of age X & no    & 0.005 & 10.01 \\
    54    & CL    & management issues & no    & 0.012 & 22.05 \\
    55    & MT    & job postings* & no    & 0.005 & 10.21 \\
    56    & CL    & need help & no    & 0.058 & 111.22 \\
    58    & CL    & entry level job issues & no    & 0.008 & 15.4 \\
    59    & CL    & looking for tips (best ways how to) & yes   & 0.004 & 8.59 \\
    62    & CL    & online jobs/ extra money & yes   & 0.051 & 91.83 \\
    64    & CL    & not meeting work experience required by job ads & no    & 0.075 & 139.67 \\
    65    & CL    & quit  & no    & 0.325 & 834.01 \\
    66    & CL    & dream job & no    & 0.001 & 2.92 \\
    67    & CL    & job titles, promotions & no    & 0.005 & 9.26 \\
    68    & CL    & follow-up emails/calls & yes   & 0.046 & 79.24 \\
    69    & CL    & blue collar/service job issues & no    & 0.002 & 4.63 \\
    70    & CL    & office jobs issues & no    & 0.001 & 2.51 \\
    71    & CL    & worries related to employer while job searching & nearly yes & 0.063 & 114.41 \\
    72    & CL    & retail job issues & nearly yes & 0.001 & 3.06 \\
    73    & CL    & quitting a new job & nearly yes & 0.14  & 274.59 \\
    74    & CL    & performance review & no    & 0.002 & 4.59 \\
    75    & CL    & issues involving boss & no    & 0.069 & 132.2 \\
    76    & CL    & issues involving a manager & no    & 0.008 & 13.96 \\
    77    & CL    & job hunting & yes   & 0.001 & 3.49 \\
    78    & CL    & university related questions & no    & 0.043 & 77.05 \\
    79    & CL    & asking for advice related to situation X & yes   & 0.014 & 25.11 \\
    80    & CL    & mental health & nearly yes & 0.064 & 113.24 \\
    81    & CL    & issues with coworkers & no    & 0.009 & 16.39 \\
    82    & MT    & tech companies* & yes   & 0.002 & 3.84 \\
    83    & CL    & resigning & no    & 0.042 & 77.94 \\
    84    & MT    & health issues* & yes   & 0.001 & 1.74 \\
    85    & CL    & issues related to contracts & yes   & 0     & 1.76 \\
    86    & CL    & applying for jobs & nearly yes & 0.048 & 87.88 \\
    88    & CL    & temp jobs & yes   & 0.008 & 15.81 \\
    89    & MT    & looking for jobs* & yes   & 0.045 & 84.63 \\
    90    & MT    & asking for advice  & yes   & 0.019 & 31.21 \\

\label{tab:si_topc_did}
\end{longtable}

\end{document}